\documentclass{aastex631}

\newcommand{\rev}[1]{\textcolor{black}{#1}}
\newcommand{\revv}[1]{\textcolor{black}{#1}}
\usepackage{amsmath}

\usepackage{array}
\newcolumntype{L}[1]{>{\raggedright\let\newline\\\arraybackslash\hspace{0pt}}m{#1}}
\newcolumntype{C}[1]{>{\centering\let\newline\\\arraybackslash\hspace{0pt}}m{#1}}
\newcolumntype{R}[1]{>{\raggedleft\let\newline\\\arraybackslash\hspace{0pt}}m{#1}}

\begin{document}

\title{An Introduction to High Contrast Differential Imaging of Exoplanets and Disks}

\author{Katherine B. Follette}
\affiliation{Amherst College, Department of Physics and Astronomy}



\begin{abstract}

This tutorial is an introduction to High-Contrast Imaging, a technique that enables astronomers to isolate light from faint planets and/or circumstellar disks that would otherwise be lost amidst the light of their host stars. Although technically challenging, high-contrast imaging allows for \textit{direct} characterization of the properties of detected circumstellar sources. The intent of the article is to provide newcomers to the field a general overview of the terminology, observational considerations, data reduction strategies, and analysis techniques high-contrast imagers employ to identify, vet, and characterize planet and disk candidates.  

\end{abstract}



\section{Introduction \label{sec:intro}}
One of the breakthrough technologies of modern exoplanet astronomy is the technique of high-contrast imaging (HCI, often referred to more simply as ``direct imaging"). HCI is a catchall term that encompasses the instrumental hardware, image processing techniques, and observing strategies that are employed to enable astronomers to image very faint sources (planets, circumstellar disks) in the vicinity of bright stars. 

This article provides a basic introduction to the challenge of high contrast imaging in Section \ref{sec:intro}. It then defines and briefly describes the hardware involved in HCI in Section \ref{sec:hardware}. In Section \ref{sec:HCIPSF}, it outlines how hardware and atmospheric aberration manifest in the anatomy of a HCI Point Spread Function (PSF). Section \ref{sec:diffim} introduces the range of ``differential imaging" observational techniques that are employed to facilitate separation of starlight from disk or planet light in post-processing, and Section \ref{sec:algos} outlines the algorithms used to do so. Section \ref{sec:analysis} describe analysis techniques commonly employed to extract the properties of imaged planets and disks from post-processed HCI images, and Section \ref{sec:falsepos} describes potential sources of false positives. Technologies that complement HCI are covered briefly in Section \ref{sec:othertech}. \rev{The article is accompanied by a python code tutorial containing sample implementations of each of the main differential imaging techniques, as well as exercises for the reader. It is available at \url{https://github.com/kfollette/PASP_HCItutorial}.}

\rev{Throughout this article, I include definitions of  many terms and phrases peculiar to High-Contrast imaging, but also assume knowledge of some common astronomy and optics terms that readers just getting started in the field may not yet be familiar with. Furthermore, the references I've chosen to include in the main text are primarily to the foundational work(s) that developed a particular technique. They are intended merely as a starting point, and should not be interpreted as the ``state of the art" in the field. I provide two living documents to accompany the tutorial that I hope will serve as references in both areas. The first (available at \url{https://bit.ly/HCIjargon}) provides definitions of key astronomy and optics jargon used throughout this tutorial, which some readers may find useful when they encounter unfamiliar terms. The second (available at \url{bit.ly/beginHCI}) provides a recommended reading and viewing list for those who would like to delve deeper into the techniques discussed here.} 

\subsection{What is high-contrast imaging?}
\begin{figure}
    \centering
    \includegraphics[width=\textwidth]{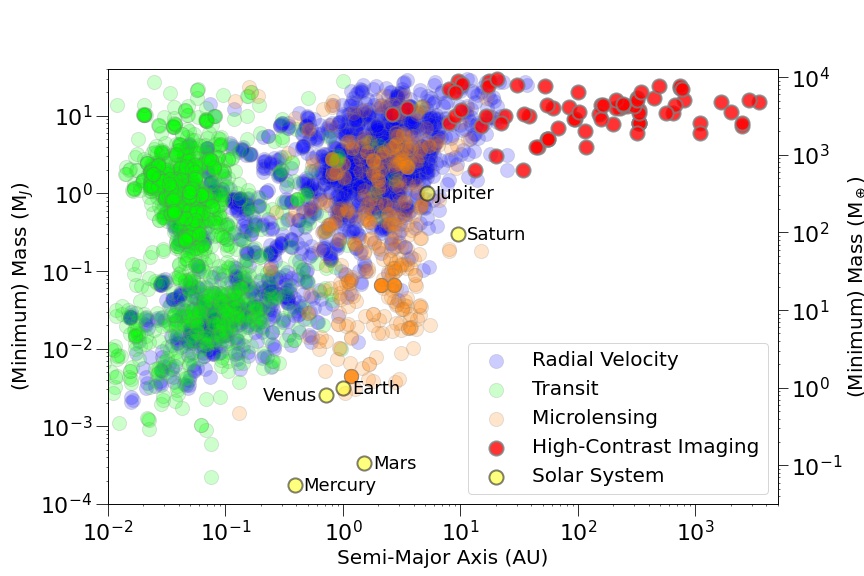}
    \caption{The population of known exoplanets discovered with high-contrast imaging (red) as compared to those found with indirect methods: transits (green), radial velocity (blue), and microlensing (orange) as of February, 2023 per the \href{https://exoplanetarchive.ipac.caltech.edu/}{NASA Exoplanet Archive}. Exoplanets are shown relative to solar system planets (yellow), highlighting the fact that detection techniques are not yet capable of detecting solar system analogs.}
    \label{fig:landscape}
\end{figure}

The High-Contrast Imaging (HCI) technique is a relative newcomer in the world of exoplanet detection techniques, with the first discoveries in 2004 and 2008 \citep{Chauvin2004, Marois2008, Kalas2008}. Although the number of planet detections is to date lower for high-contrast imaging  \footnote{As of the writing of this tutorial, $\sim$50 companions have been imaged with estimated masses below the canonical ``deuterium-burning" limit of $<$13M$_J$ (the formal boundary between ``planet" and ``brown dwarf", though the utility of this boundary as a defining line between populations is debatable). However, this number more than doubles when considering all bound substellar ($<$70M$_J$) companions to higher mass stars. Brown dwarf companions with masses less than $\sim$20$M_{Jup}$ are often referred to as ``Planetary Mass Companions" (PMCs), and are likely part of the same underlying population as (i.e. formed similarly to) many of the objects currently classified as directly imaged ``planets" \citep{Wagner2019}.} than for the indirect (radial velocity, transit, and microlensing) techniques, directly imaged companions are arguably the best characterized exoplanets. HCI also provides the best prospects for current and future characterization of exoplanet atmospheres, particularly temperate ones conducive to life as we know it. The commitment of the community to this goal is evident in the first theme of \textit{Pathways to Discovery in Astronomy and Astrophysics for the 2020s} (also known as the Astro2020 Decadal Survey) -- ``Pathways to Habitable Worlds".  It calls for a "step-by-step program to identify and characterize Earth-like extrasolar planets, with the ultimate goal of obtaining \textbf{\textit{imaging}} and \textbf{\textit{spectroscopy}} of potentially habitable worlds" \citep[pg. 2,][emphasis mine]{Astro2020}. \rev{The gap between the modern directly imaged planet population and Earth-analogs is large in both mass and semi-major axis space (see Fig. \ref{fig:landscape}). However, while indirect planet detection methods are currently more sensitive to terrestrial planets, the decadal survey goal of \textbf{\textit{imaging}} and \textbf{\textit{spectroscopy}} of exo-Earths cannot be achieved without direct detection.}

Although the current state of the art in HCI is imaging of $>$1M$_J$ planets at $\sim$ tens of au separations, the future of the technique is bright (pun intended!), and vigorous ongoing technology development will push its sensitivities to lower mass and more tightly-separated planets. 

\subsection{What is Contrast?}
In the context of HCI, the term ``contrast" refers to the brightness ratio between an astronomical source (planet, disk) and the star it orbits. ``High" contrast images are those where the ratio $\frac{F_{source}}{F_{star}}$ is small, meaning the source is much fainter than the star -- these detections are difficult. ``Low" contrast images are therefore ones where the source-to-star ratio is larger, meaning the source is brighter relative to the star -- these detections are less challenging. 

   
\begin{figure}
    \centering
    \includegraphics[width=0.75\textwidth]{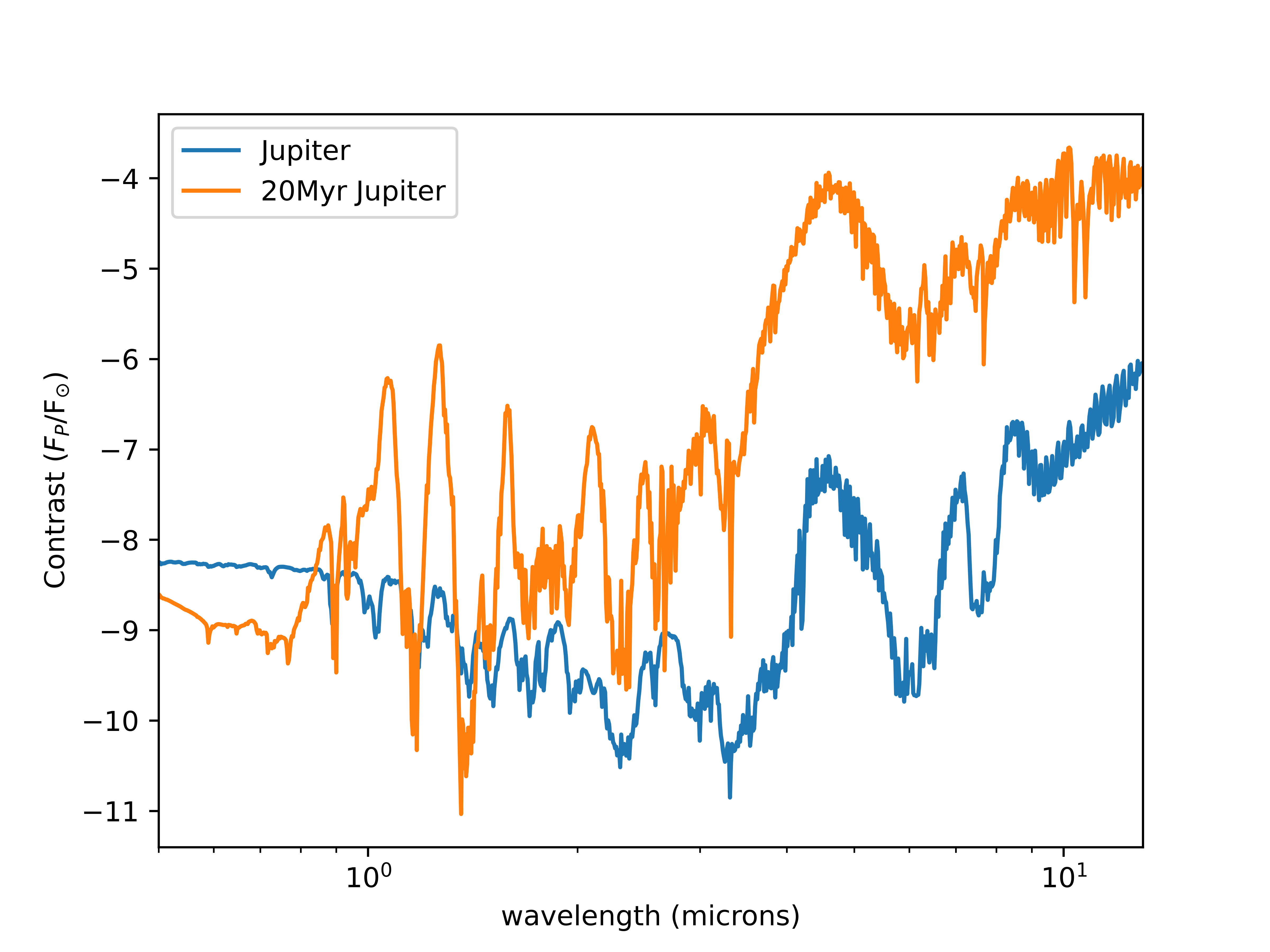}
    \caption{The predicted contrast ratios required to image Jupiter both as an ``old" (4.5Gyr, blue) and ``young" (20Myr, yellow) planet as a function of wavelength. Thermal and reflected light spectra were generated for both planets with PICASO \citep{Batalha2019}, binned to a spectral resolution of 300, and summed. The young Jupiter's atmospheric properties were generated using the SONORA cloud-free atmospheric model grid \citep{Marley2021} and divided by a simulated spectrum for a star with properties appropriate for the young Sun \citep[T=4300K, logg=4.3, R=1.2R$_{\odot}$][]{Baraffe2015}. The ``old" Jupiter spectrum was generated for a 90\% cloudy/10\% cloud-free surface and divided by a solar spectrum.} 
    \label{fig:jupiter}
\end{figure}

Unlike stars, where absolute brightness is almost entirely a function of mass, for planets, brightness is a function of both mass and age. Planets begin their lives hot and \rev{bright and}, lacking an internal source of energy sufficient to maintain that temperature, cool with time.

\rev{As they evolve, planetary spectra, and therefore contrast, also change drastically. Figure \ref{fig:jupiter} shows contrast at a range of wavelengths for the same planet (Jupiter) when ``young" (20Myr) and ``old" (4.5Gyr, the age of our Solar System). It highlights the extreme variation in contrast as a function of wavelength as planets age}. 

In thinking about contrast for point sources, it is useful to keep several benchmark quantities in mind, namely:
\begin{itemize}
    \item In the near-infrared (1-3$\mu$m), young ($\sim$few to few tens of Myr) giant planets generally have contrasts in the range $\sim10^{-5}--10^{-6}$ relative to their host stars. They radiate away much of their initial thermal energy over the course of the first tens of millions of years after formation, thus higher contrasts are required to detect them as they get older.
    \item At 3-5$\mu$m, the same young ($\sim$few Myr) planets, have more moderate contrasts of $\sim10^{-3}--10^{-4}$.  With temperatures of $\sim$500-1500K, this is because their thermal emission peaks in this wavelength regime, and the brightness gap relative to the much brighter and hotter (peak emission bluer) star is narrowed. \rev{This remains the region of most favorable contrast even as planets age}. 
    \item In the optical, planets have undetectably low levels of direct thermal emission, and are seen instead in reflected light (stellar photons redirected/scattered by their atmospheres toward Earth). For mature planets ($\gtrsim$100Myr), this wavelength regime provides \rev{more moderate contrasts than the NIR. For example, at 4.5Gyr, Jupiter and Earth have contrasts of $\sim10^{-9}$ and $10^{-10}$, respectively at 0.5$\mu$m. Combined with resolution advantages inherent in shorter wavelength imaging (See Section \ref{sec:hardware} for details) optical wavelengths provide the best prospects for future detection of solar system analog planets.}
\end{itemize}

\begin{figure}
    \centering
    \includegraphics[width=0.75\textwidth]{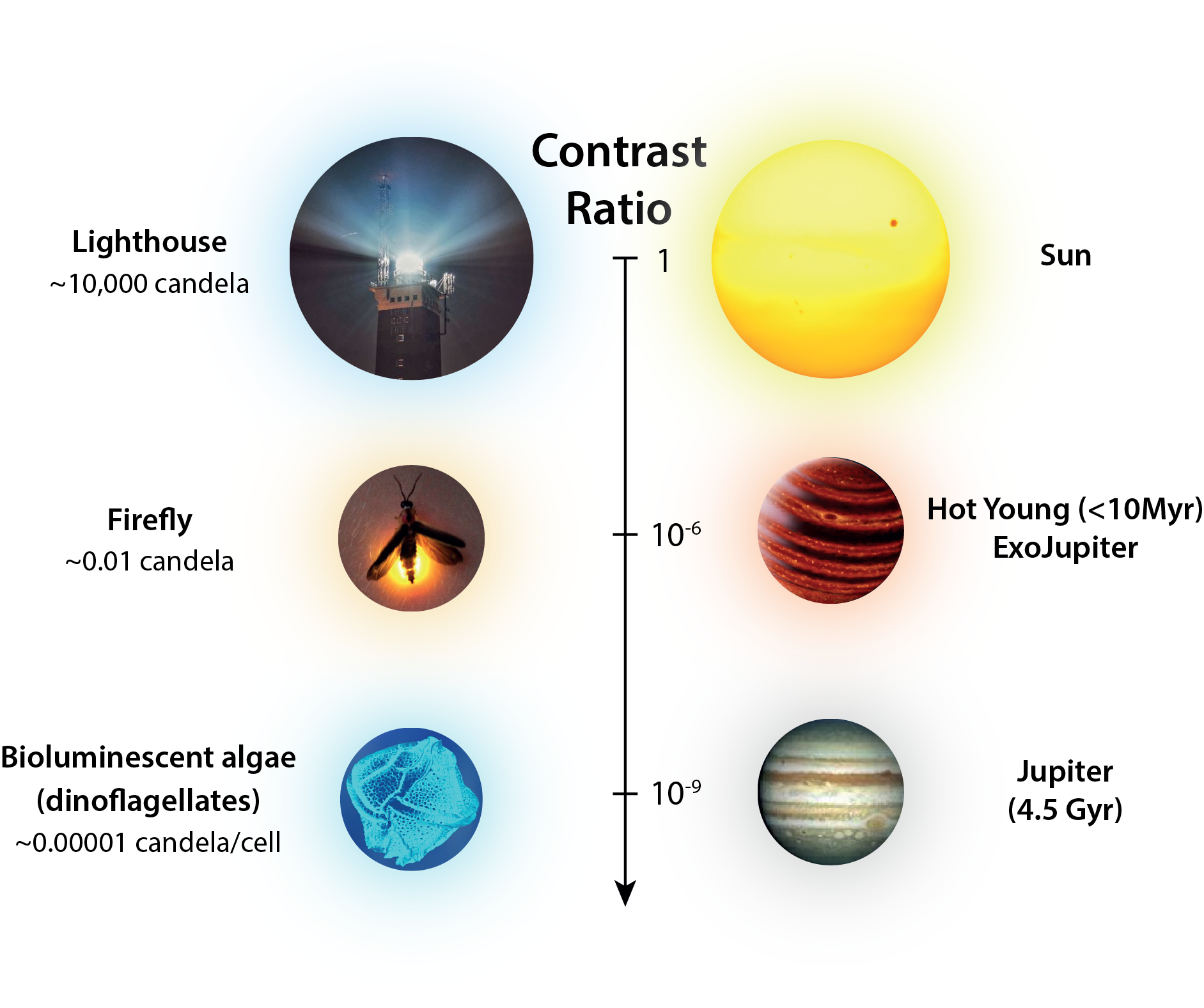}
    \caption{A schematic illustration of the magnitude of the brightness differential between the sun and a hot, young exojupiter in the NIR and the sun and a reflected light Jupiter in the optical. The brightness differential for a young Jupiter analog is $\sim$10$^{-6}$, comparable to the brightness differential between a lighthouse and a firefly. Once a jupiter-like planet has radiated most of the energy of formation and no longer glows brightly in the infrared, this differential drops to 10$^{-9}$, akin to the brightness differential between a lighthouse and a single bioluminescent alga cell.}
    \label{fig:lighthouse}
\end{figure}

\rev{A simple analogy will help drive home the near (but not wholly) intractable nature of the contrast problem}. As shown in Figure \ref{fig:lighthouse}, for thermal emission from hot young exojupiters, the contrasts outlined above are comparable to the ratio of light emitted by a firefly relative to a lighthouse. For true (4.5 Gyr) Jupiter analogs in optical reflected light, a more apt comparison is a single bioluminescent alga relative to a lighthouse. This highlights the tremendous technological barriers that the field must overcome in order to achieve direct characterization of mature, potentially-habitable exoplanets. 

Precisely how hot a planet is at formation (and therefore how bright it appears) depends on how it was formed, and a range of formation modes are likely to overlap within the exoplanet population. In other words, planets (and brown dwarfs) of the same mass may have formed via different mechanisms.  

Planets like those in our solar system most likely formed via a ``cold start" mechanism involving the gradual assembly of solid material within a circumstellar disk. Their ``cold" starts are only cold in comparison to so-called ``hot start" planets, which also form in a circumstellar disk, but rapidly as a result of gravitational \rev{collapse}. 
The high masses and wide separations of most directly imaged planets make them good candidates for hot start formation, but current and next-generation instruments are detecting lower mass, closer-in planets for which formation mechanism is more ambiguous, and could proceed under either path. The range of models and their predictions and assumptions is well-described in \citet{Spiegel2012}. For our purposes, the most important takeaways are that directly imaged exoplanet brightnesses can only be translated to mass estimates under assumptions of: (a) stellar age, and (b) planetary formation pathway/initial entropy of the planet \rev{unless a direct measure of the planet's mass is available from another method, such as astrometry or radial velocity}.

\subsubsection{What do we learn from HCI planet detections?}
The simplest measurements made for individual directly-detected exoplanets are their \textbf{locations}\footnote{In this section, I will place observed properties in \textbf{bold} the first time I reference them, and inferred physical properties in \textbf{\textit{bolded italics}}} (astrometry) and \textbf{brightnesses} (photometry). Together with {\textit{evolutionary models} for young giant planets (which assume a formation pathway, e.g., \citealt{Baraffe2003}), \textbf{photometric data} allow for inference of a planet's \textbf{\textit{mass}}, provided the system has a well-constrained  \textbf{distance}\footnote{Nearly all HCI detections are for objects in the solar neighborhood, for which \textit{Gaia} distances are sufficiently robust to consider them directly measured, rather than inferred quantities. For non-parallax distance measurements, this is not necessarily true.} \rev{and a moderately-constrained \textbf{\textit{age}}}. 

Given the difficulty of robustly estimating ages for young objects, the preferred targets for direct imaging surveys have been young moving group stars; age estimates for these coeval groups are better constrained by averaging across independent estimates for their many members. \textbf{Planetary luminosity} and age can also be compared to the predictions of various planet formation models \citep[e.g. the so-called cold/warm/hot start models, ][]{Spiegel2012} to inform the initial conditions under which planets are born.

The combination of \textbf{detection limits} of large HCI planet-finding campaigns and evolutionary models allows for constraints on the \textbf{\textit{occurrence rates}}  of populations of exoplanets in various mass and separation ranges unique to direct imaging (currently $\gtrsim$1M$_J$ and $\gtrsim$10au). Population constraints, in turn, inform formation models. For a review of what was learned about planet populations from the first generation of HCI campaigns, see \citet{Bowler2016}.

\textbf{Orbital monitoring} of directly imaged planets also provides constraints on the dynamical evolution of young planetary systems. For example, \textbf{\textit{coplanarity}} and the prevalence of \textbf{\textit{orbital resonances}} in multi-planet systems inform planet formation and migration models \citep[e.g.][]{Konopacky2019}. \textbf{\textit{Alignment}} (or misalignment) of planetary orbits with the stellar spin axis and/or the circumstellar disk plane informs the history of dynamical interactions within the system \citep[e.g.][]{Balmer2022, Brandt2021}. Similarly, dynamical characterization of planets in systems with disk features hypothesized to be planet-induced provides a means to test disk-planet interaction models \citep[e.g.][]{Fehr2022}. For a comprehensive review of planetary dynamical processes, see \citet{Davies2014} and \citet{Winn2015}.

Finally, \textbf{spectroscopy} of imaged companions allows for direct characterization of \textbf{\textit{atmospheric properties}}. To first order, low resolution spectra can inform the bulk \textbf{\textit{composition}} of the atmosphere in more detail than photometry alone. For instance, even a low-resolution infrared spectrum of a giant planet can inform whether its atmosphere is CH$_4$ or CO-dominated. Directly imaged planet spectra, in combination with detailed atmospheric models, can also inform the \textbf{\textit{temperature-pressure structure}} of the atmosphere, likely \textbf{\textit{condensate (cloud) species}}, and even the prevalence of photo- and disequilibrium \textbf{\textit{chemical processes}}. Constraints on \textbf{\textit{C/O ratios}} of planetary atmospheres are probes of their formation locations relative to various ice lines that determine whether these elements are found in the gas or solid phase. 

The advent of medium resolution spectroscopy of directly imaged planets with instruments such as VLT GRAVITY (R$\sim$500 in medium resolution mode) is enabling stronger constraints on these properties, with upgrades planned at the VLT to improve resolutions even further. Very high-resolution spectra of directly imaged companions will be enabled by coupling focal-plane optical fibers to existing high-resolution (R$\sim$30,000) spectrographs \citep[e.g. The Keck Planet Imager and Characterizer (KPIC),][]{Mawet2016}. Such work requires very precise knowledge of planet astrometry to enable fiber placement, but will enable very exciting science such as constraints on planetary \textbf{\textit{rotation rates}}, which can be compared against the predictions of various formation models. For a review of spectroscopy of directly imaged planets, see \citet{Biller2018} and \citet{Marley2007}.

\subsubsection{What do we learn from HCI disk detections?}
HCI's detection efficiency is significantly higher for cicumstellar disk structures than for planets\rev{\footnote{The higher efficiency of disk detections is in part because of their ubiquity at detectable radii around young stars, but also because disks, unlike most planets, are detectable in polarized light. Polarized light from disks can be efficiently separated from unpolarized starlight via Polarized Differential Imaging (see Section \ref{sec:PDI}).}}, and many high-resolution high-contrast images of circumstellar material have been collected by exoplanet direct imaging surveys \citep[e.g.][]{Rich2022,Esposito2020,Avenhaus2018}. Such observations provide direct constraints on the distribution and composition of planet-forming material. Symmetric morphological features (such as rings, gaps, and cavities), inform the distribution of dust in planet-forming systems and, likely, the architectures of their planetary systems. Asymmetric features (such as warps and spiral arms) provide indirect evidence of embedded or undetected planetary perturbers and/or likely locations for future planet formation. These ``signposts" of planet formation, though difficult to interpret, provide a wealth of information about planets and planet formation \textit{at or near the epoch of formation}. For a comprehensive review of the state of high-contrast disk imaging, see \citet{Benisty2022}.

NIR HCI disk images are also extremely powerful in combination with high-resolution millimeter imagery. In the millimeter and sub-millimeter, dust continuum emission traces large grains in the disk midplane, and millimeter line emission can be used to trace various gas-phase species as well. NIR high-contrast images trace an entirely different population, namely small micron-sized dust grains in the upper layers of the disk. Thus, the combination of NIR and mm high-resolution imagery yields a holistic picture of various disk components, a powerful combination for understanding the radial and vertical structure of disks. 

Finally, multiwavelength NIR high-contrast imagery can be used to constrain grain properties such as size, porosity, and composition \citep[e.g.][]{Chen2020}, as well as the water ice content of NIR-scattering grains \citep[e.g.][]{Betti2022}. A good understanding of grain properties is essential to understanding the microphysics of the dust coagulation that will eventually form planets. 

\section{Enabling Technologies for High-Contrast Imaging \label{sec:hardware}}

HCI is built upon a foundation of enabling technologies, namely: adaptive optics, coronagraphy, wavefront sensing, and differential imaging techniques, each of which is introduced in this section.  For a more comprehensive technical review of many of these technologies, see \citet{Guyon2018}.

\subsection{Adaptive Optics}

Adaptive optics is perhaps the most critical HCI enabling technology for ground-based imaging campaigns. Without it, image resolutions are limited by astronomical seeing, or the size of coherent patches in the earth's atmosphere (approximated by the ``Fried parameter" $r_0$, which has a $\lambda^{6/5}$ dependence). With adaptive optics, modern HCI instruments can approach the diffraction limit, 
$$\theta = 1.22\frac{\lambda}{D}$$
where $\lambda$ is the wavelength and D the diameter of the telescope. Table \ref{tab:res} gives the diffraction-limited resolution of an 8m telescope at 0.55$\mu$m (V band), 1.6 $\mu$m (H band) and 3.5$\mu$m (L band) in physical units as compared to the seeing limit at an exceptional telescope site under good weather conditions (0$\farcs$25 at 0.55$\mu$m) at each wavelength.

In principle, the diffraction-limited Point Spread Function (PSF)\footnote{A Point Spread Function describes the appearance of a point source when imaged with a given combination of telescope, instrument, and wavelength. In functional form, it describes the location and intensity of light across the image plane.} of a circular telescope aperture is the ``Airy pattern". In practical terms, the function describing this PSF places the majority of the incoming starlight into a ``diffraction-limited core", with a radius of $1.22\lambda/D$ and a Full Width at Half Maximum (FWHM) of $1.03\lambda/D$.  Extending from this central core are a characteristic set of ``Airy" diffraction rings that decrease in amplitude outward and are spaced by roughly $1\lambda/D$ from one another with the first minimum at $1.22\lambda/D$. In a perfect diffraction-limited system, the central ``Airy disk" contains 84\% of the total light in the PSF, with the remainder of the light in the Airy rings. 

\begin{table}[]
    \centering
    \begin{tabular}{cccc}
        \textbf{Distance} &  \multicolumn{3}{c}{\textbf{Resolution (in au)}} \\
        (pc) &  @$0.55\mu m$ & @$1.65\mu m$ & @$3.5\mu m$ \\
        \hline\hline
        \multicolumn{4}{c}{Seeing-Limited Observations}\\
        \hline
        50 & 12.5 & 46.5 & 115 \\
        150 & 37.5 & 140 & 345 \\
        \hline
        \multicolumn{4}{c}{Diffraction-Limited Observations}\\
        \hline
        50 & 0.9 & 2.6 & 5.5 \\
        150 & 2.6 & 7.8 & 16.5 \\
        \hline
    \end{tabular}
    \caption{Seeing ($r_0$) and diffraction ($\theta$)-limited resolutions at three common HCI wavelengths for an 8m telescope at an excellent astronomical site in good weather conditions (0$\farcs$25 seeing at V band). Values are given in astronomical units for objects at distances of 50pc (the volume limit of many HCI surveys) and 150pc (a typical distance to nearby star forming regions).}
    \label{tab:res}
\end{table}

\begin{figure}
    \centering
    \includegraphics[width=\textwidth]{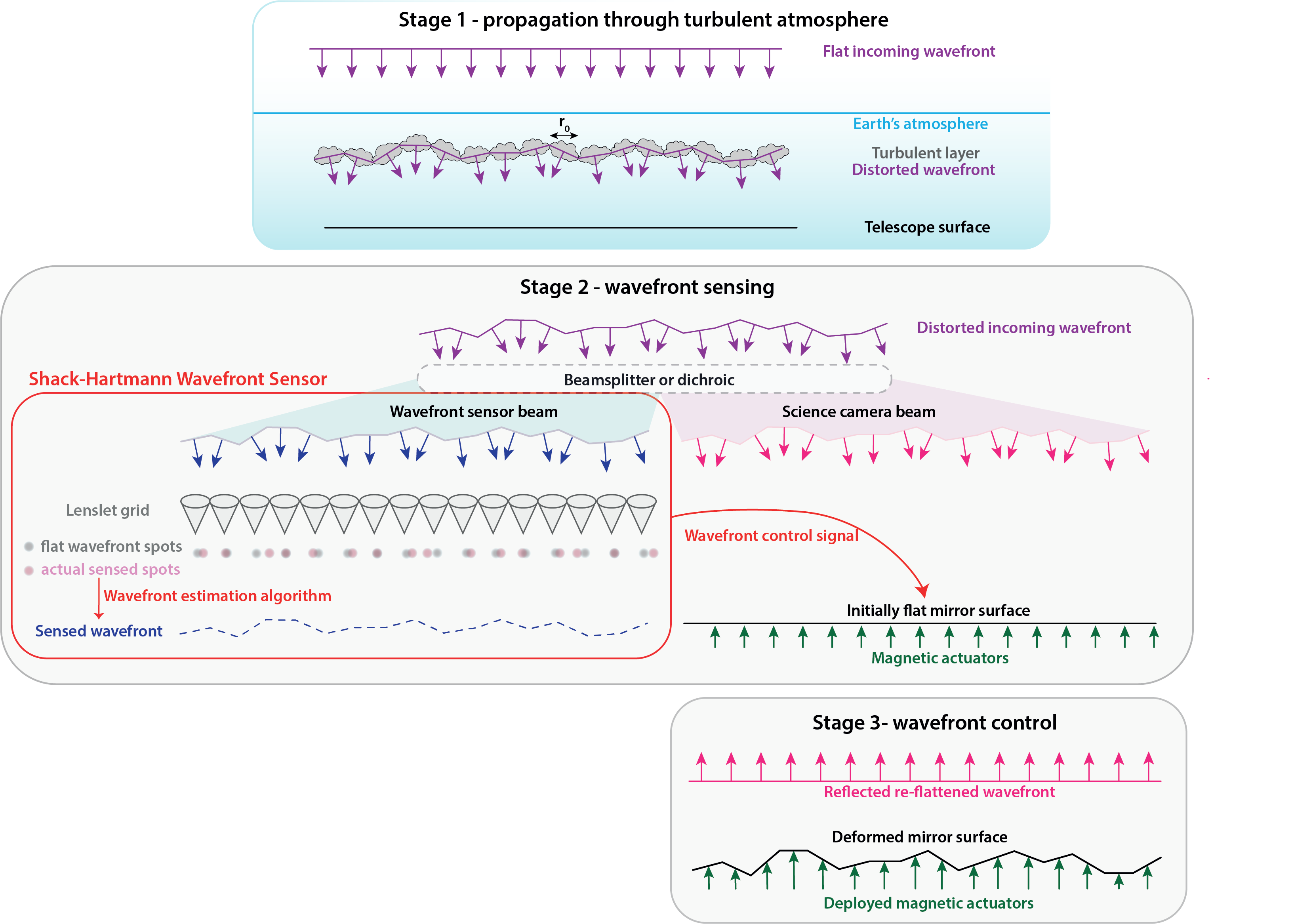}
    \caption{A simplified, schematic illustration of the process of adaptive optics. \textbf{\textit{``Stage 1"}} depicts the effect of the Earth's atmosphere on incoming plane-parallel light. The wavefront is aberrated inside of locally coherent patches in the atmosphere, and enters the telescope aperture with corrugations of a characteristic size ($r_0$). In \textbf{\textit{``Stage 2}}, the incoming light is passed through a beamsplitter or dichroic, which splits it, sending some to a wavefront sensor and the rest to a science camera. In this case, a Shack-Hartmann wavefront sensor (see Section \ref{sec:wfs}) is depicted, wherein an array of lenslets is inserted into the focal plane. Each makes a spot whose location relative to the orientation of the lenslet is indicative of the slope of the incoming wavefront. The spot locations are converted to a ``best guess" of the incoming wavefront shape and a corresponding control signal is sent to actuators under an (initially flat, generally tertiary) mirror. \textbf{\textit{``Stage 3"}} depicts the result of the deformed wavefront reflecting off of the deformed mirror, causing the reflected wavefront to be re-``flattened", thus compensating for atmospheric aberration. The sensed wavefront is depicted here as an unrealistically perfect match to the true incoming wavefront. In reality, kHz-scale time variation in the incoming wavefront, unsensed or imperfectly estimated wavefront aberration, and the speed and nature of the control algorithm mean that no wavefront is perfectly sensed and corrected. Some residual corrugation will always remain in a real AO system.}
    \label{fig:aoschem}
\end{figure}

\rev{In the case of a telescope with a circular aperture and a central obscuration (e.g. by a telescope secondary mirror)} the Airy pattern has a functional form of: 
$$I(u)= \dfrac{1}{(1-\epsilon^2)^2}\Bigg[\dfrac{2J_1(u)}{u}-\epsilon^2\dfrac{2J_1(\epsilon u)}{\epsilon u}\Bigg]^2$$
where u is a dimensionless radial focal plane coordinate defined as:
$$u=\dfrac{\pi}{\lambda}D\theta$$
and $\theta$ is defined as the angle between the optical axis and the point of observation. The center of the PSF is at $\theta$=0 and therefore u=0, and I(u) is the PSF intensity at location u. The quantity $\epsilon$ is a measure of the amount of central obscuration expressed as a fraction of the total aperture \rev{(which acts to decrease the effective aperture and thus the predicted peak intensity)}, and J$_1$ is the first order Bessel function of the first kind.

In practice, HCI \rev{PSFs} tend to be dominated by Airy or Airy-like diffraction patterns with a few key deviations. First, no modern AO systems achieve perfectly diffraction-limited performance. The PSF of a modern adaptive optics PSF is often characterized by its so-called ``Strehl Ratio" (SR), which is the ratio of \rev{a star's observed peak intensity relative to that of its theoretical diffraction-limited peak intensity. \footnote{This theoretical PSF is not fully approximated by the relatively simple $I_{u=0}$ described above for a given telescope aperture size (D), central obscuration $\epsilon$, and wavelength $\lambda$, because (a) it assumes no other obscurations in the aperture (e.g. secondary mirror supports, downstream optical elements), and (b) it computes the PSF for a single wavelength, which is not measurable in practice. Thus, real Strehl Ratio approximations require detailed instrumental PSF models that include all of the telescope and instrument system's optical elements. The on-sky predicted PSFs are then normalized to the same total intensity and divided to approximate Strehl Ratio. For further discussion of the subtleties of Strehl Ratio determination, see \citet{Roberts2004}.}}. Modern Extreme Adaptive Optics (ExAO) systems routinely achieve SRs of 80-95\% in the Near Infrared, but only 10-30\% in the optical at present. 


A proper treatment of the effect of the atmosphere on incoming starlight requires detailed atmospheric turbulence modeling (e.g. a Kolmogorov model). However, a decent first-order approximation of the effect of the Earth's atmosphere on incoming starlight, depicted in Figure \ref{fig:aoschem}, is to imagine a plane-parallel electromagnetic wave\footnote{Plane-parallel here means that if we were to draw a shape connecting equivalent phases of incoming electromagnetic waves from the same source, say the location where their electric field strengths are strongest, the shape of our equal-phase surface would be a plane perpendicular to the direction of travel. In other words, light from a distant source enters the Earth's upper atmosphere in phase with neighboring light waves. This is an approximation because light exits a spherical object symmetrically, meaning that a surface of constant phase should always have some curvature; however, the distances to astronomical objects are vast compared to the sizes of the telescopes we use to intercept their light. This means that we intercept only a tiny area of a vast spherical shell of light from the star, a shell so vast that the tiny area we intercept can be treated as locally ``flat".} with some constant phase and amplitude encountering a layer in the Earth's atmosphere composed of coherent patches of size $r_0$ (atmospheric ``cells"). Inside these cells, the wavefront phase is aberrated such that it remains locally flat, however phase offsets occur between neighboring cells. Phase aberrations can take many forms and are often represented as an orthogonal basis set of polynomials with both radial and azimuthal dependencies (e.g. the Zernike polynomials). Low order aberrations have familiar names, and ones that you're likely to encounter in your annual eye exam, such as ``astigmatism" and ``coma". Higher order aberrations take more complex forms in phase space, but all are essentially disruptions in the intrinsic shape of the incoming PSF. For illustrative purposes, let's imagine only the simplest two low-order modes, the so-called ``tip" and ``tilt" modes, which preserve the shape of the PSF but modify the direction of the incoming wavefront relative to the original travel direction.

The effect of tip/tilt aberrations is that a wavefront exiting a layer of atmospheric cells is no longer plane-parallel. Instead, it is corrugated (the angle of arrival varies across the telescope aperture, see Figure \ref{fig:aoschem}'s ``distorted incoming wavefront") with some wavelength-dependent characteristic length scale (The Fried coherence length, $r_0\sim\lambda^{6/5}$). For an atmospheric layer at a certain height in the atmosphere, this characteristic length scale can also be represented as a characteristic angular scale called the ``isoplanatic angle", $\theta_0$. Note again that this is just a first-order approximation, albeit a useful one for building intuition, and that, in reality, there are a number of aberrating layers in the atmosphere with their own characteristic coherence lengths, heights, and wind speeds. The practical consequence when integrated over the telescope aperture is that the light of each coherent patch manifests as its own diffraction limited PSF at a different location in the image plane centered around the optical axis of the telescope. The instantaneous result is a number of superposed independent images of the star equal to the number of coherent atmospheric patches that the wavefront incident on the telescope passed through - i.e. the image is blurry. 

Locally-coherent patches at a given layer in the atmosphere only remain so on timescales of hundredths- to thousandths- of a second (due to wind, temperature/pressure variation, etc.) which means Adaptive Optics systems must operate on these timescales in order to detect and correct  these aberrations with Wavefront Sensors (WFS). Let's extend our toy example of an incoming plane-parallel wavefront that experiences pure tip/tilt aberrations at a single layer in the atmosphere. \rev{Imagine a series of corrugated wavefronts exiting this layer and being collected continuously by an astronomical detector} over a realistic exposure time of several to several tens of seconds. The result will be a superposition of many hundreds or thousands of diffraction-limited PSFs (so-called ``speckles") at various locations relative to the central optical axis. The result is a seeing-limited PSF, whose size/FWHM will vary according to various properties of the atmosphere, but will always be much larger than the diffraction limit. Modern AO systems are able to operate at 1-2kHz frequencies, however they are not able to perfectly sense the wavefront nor to perfectly or completely correct it on the relevant timescales. \rev{Many advancements are being made in both the hardware and software of wavefront control, including the advent of algorithms that attempt to account for the time delay between sensing and applying a wavefront correction by predicting the state of the wavefront into the future \citep[so-called ``predictive control" algorithms, e.g.][]{Poyneer2007,Guyon2017}.}

The consequence of a perfect AO system that could fully detect for and correct wavefront aberration would be a perfect SR=100\% diffraction-limited PSF. The reality is of course not perfect - a partially- or imperfectly-corrected wavefront results in the alignment of many but not all of these instantaneous PSFs. Some uncorrected, residual seeing-limited ``halo" with a width of approximately $\frac{\lambda}{r_0}$ is expected, and it's amplitude should decrease as the performance of the AO system (Strehl Ratio) improves. 
Imperfect wavefront correction can also lead to certain persistent speckles, so-called ``quasi-static speckles", that are stable on timescales of minutes to hours. These are particularly worrisome because they can mimic planets, but they have the advantage of being static in their location in the instrument frame. They also exhibit spectra that are identical to that of the central star. These properties make them amenable to removal by angular and spectral differential imaging (ADI/SDI, see Section \ref{sec:diffim}). 

NIR HCIs can have a dozen or more clear, detectable Airy rings in their unocculted AO PSFs. These Airy rings present a fundamental barrier to achieving high contrast in the environs of the central star, and additional optics are often employed to mitigate them. Because Airy rings are a consequence of diffraction at the edges of the entrance pupil, mitigating optics are generally pupil plane\footnote{Complex modern instruments utilize optics in both the ``image plane", where light incident on the telescope is brought to a focus, and the ``pupil plane", where light is collimated. Estimates of the appearance of an object in a given plane can be accomplished by Fourier transform of its appearance in the other. While image plane images show the on-sky source (often manipulated by upstream optics such as coronagraphs), pupil plane images are essentially images of the entrance aperture (which you can prove to yourself with a simple ray-tracing diagram), containing e.g. the central obscuration from the secondary mirror, the spider arms suspending it, etc. HCI instruments, especially those that require precise placement of optical elements in the pupil plane, are often equipped with "pupil-viewing" cameras, which image this entrance aperture.} optics that block light near its edges. One example is the "Lyot stop".

The Airy PSF is also predicated on the assumption of a circular entrance aperture, which no realistic telescope entrance pupil is able to achieve. The presence of various optics, especially the secondary mirror and its supports, induce deviations from a perfect Airy PSF. To simulate an HCI PSF, therefore, requires a model of the telescope entrance aperture and any additional optics in the telescope beam. Example PSFs for a range of modern high-contrast imaging instruments are provided in Figure \ref{fig:hcianatomy}.

\subsection{Wavefront Sensing and Control \label{sec:wfs}}
In addition to deformable mirrors (DMs), adaptive optics systems require instrumentation that can sense atmospheric aberrations and convert them to DM control signals on kHz frequencies. From an observer's perspective, the most important features of this ``Wavefront Sensor" (WFS) and its accompanying control algorithm are its: wavelength, limiting magnitude, stability, and cadence. 

\paragraph{WFS wavelength} WFS operate most often at optical wavelengths. Since most HCI is done in the NIR, such systems implement a dichroic that sends all optical light to the wavefront sensor and all NIR light to the science camera. Although this results in no loss of light at the science wavelength, it does introduce a difference in the scale of the wavefront aberrations that are sensed vs. detected (namely, $(\frac{\lambda_{sensed}}{\lambda_{detected}})^{6/5}$). NIR wavefront sensing is an active area of development in HCI instrumentation for this reason. For a visible light HCI instrument, wavefront sensing in the optical generally requires a beamsplitter that results in a substantial loss of signal to the science camera (50\% or more) as light at the science wavelength is diverted to the WFS. 

\paragraph{WFS limiting magnitude} is a measure of the faintest targets for which the wavefront can be sensed, and is determined at the most basic level by the architecture of the WFS. Though there are many types of WFS, the most common are the Shack-Hartmann and Pyramid WFS. \rev{Tradeoffs in WFS qualities, such as sensitivity to wavefront errors of various scales and linearity between WFS measurements and DM commands, determine the choice of WFS architecture (for a full discussion, see \citet{Guyon2018}). From the perspective of the observer, one practical consequence of WFS architecture is the range of magnitudes for which AO correction can be accomplished.} A Shack-Hartmann WFS (SHWFS, see Figure \ref{fig:aoschem} for a simple depiction) relies on a grid of lenslets placed in the pupil plane, each of which creates a spot on the WFS camera. The location of the spot created by each lenslet is controlled by the direction of the incoming wavefront, and this shape can then be applied to the DM to correct aberrations. The limiting magnitude of a SHWFS is a fixed quantity determined by the required brightness for an individual lenslet spot to be sensed. Because the lenslets are physical optics, this cannot be modified without swapping out the grid of lenslets. A pyramid WFS, on the other hand, modulates the incoming light beam around the tip of a four-faced glass pyramid, each facet of which creates an image of the telescope pupil on a WFS camera. These four pupil images can be analyzed to reconstruct the incoming wavefront. A pyramid WFS camera's pixels can also be binned to achieve correction on fainter guide stars. Although wavefront information is lost in the binning process and the quality of the AO correction is therefore necessarily compromised, this does preserve the ability to apply (more modest) AO correction to fainter stars.

\paragraph{WFS stability} is effectively a measure of how long and under what conditions a WFS can provide continuous adaptive optics correction. When AO systems are operating in ``closed loop" mode, meaning corrections are being applied in real time, the loop will ``open" in order to protect the DM if the sensed wavefront deformations require corrections whose amplitudes are too great for the range of the DM. This is called a ``breaking" of the AO control loop. One of the more critical aspects of a wavefront control algorithm is the ``gain" applied to each sensed aberration. Gain can be thought of as a multiplicative factor applied to the sensed wavefront such that all of the sensed aberration is not corrected for at once, but instead some proportion of it. This is to avoid overcorrecting an aberration and driving the mirror into an oscillation, but also to allow more wiggle room for unsensed or incorrectly-sensed aberrations to pass by without breaking the loop. Different sensed wavefront aberrations (e.g. 'low order' and 'high order' modes) can have different gains, and this is one of the principal quantities that can be adjusted in real time during AO observations. Gain, wavefront stability, WFS signal strength, and the nature of the control algorithm all conspire to determine the stability of the AO loop - basically its ability to remain closed during an observing sequence.  

\paragraph{WFS cadence} is the timescale on which the wavefront is sensed, and is the final factor controlling the quality and stability of AO correction. In this case, the wavelength of observation and nature of the telescope site (seeing, wind speed, etc.) sets the timescale on which the incoming wavefronts change, and the AO system must run faster than this timescale in order to apply quality correction. Many/most current AO systems operate at 1-2kHz frequencies, with faster speeds being required at shorter wavelengths. 

\subsection{Coronagraphy}
Another enabling HCI technology is coronagraphy, which utilizes one or more physical optics inside the instrument system to suppress direct and diffracted starlight before it reaches the detector. \rev{This allows for the collection of deeper images of planetary systems, as longer integration times can be used before saturation of the primary star.} Coronagraphy is distinct from external occulters (``starshades") and software algorithms (``wavefront control") that are designed to do similar things.   Available coronagraphic architectures have been rapidly expanding in recent years, and I will not provide a comprehensive review here, but will instead focus on the practical effects of a coronagraph for image processing. 

The purpose of a coronagraph is to redirect starlight away from the image plane by blocking or modulating it with one or more optical components, thus reducing the amount of light that must later be removed in post-processing in order to image faint companions. 

Coronagraph optical components can modulate wavefront amplitude or phase or, in many cases, both. The most basic coronagraphic architecture is an opaque or reflecting image plane spot in the center of the field, which prevents on-axis light from the central star from reaching the detector. Other coronagraphic architectures utilize interferometric techniques (e.g. the ``vortex" coronagraph) to accomplish the same goal. Additional optics are often placed in the pupil plane to mitigate diffraction around coronagraph edges and around the edges of the entrance aperture more generally, which effectively decreases the amplitude of the Airy rings and allows for higher contrast imaging. 




\begin{figure}
    \centering
    \includegraphics[width=\textwidth]{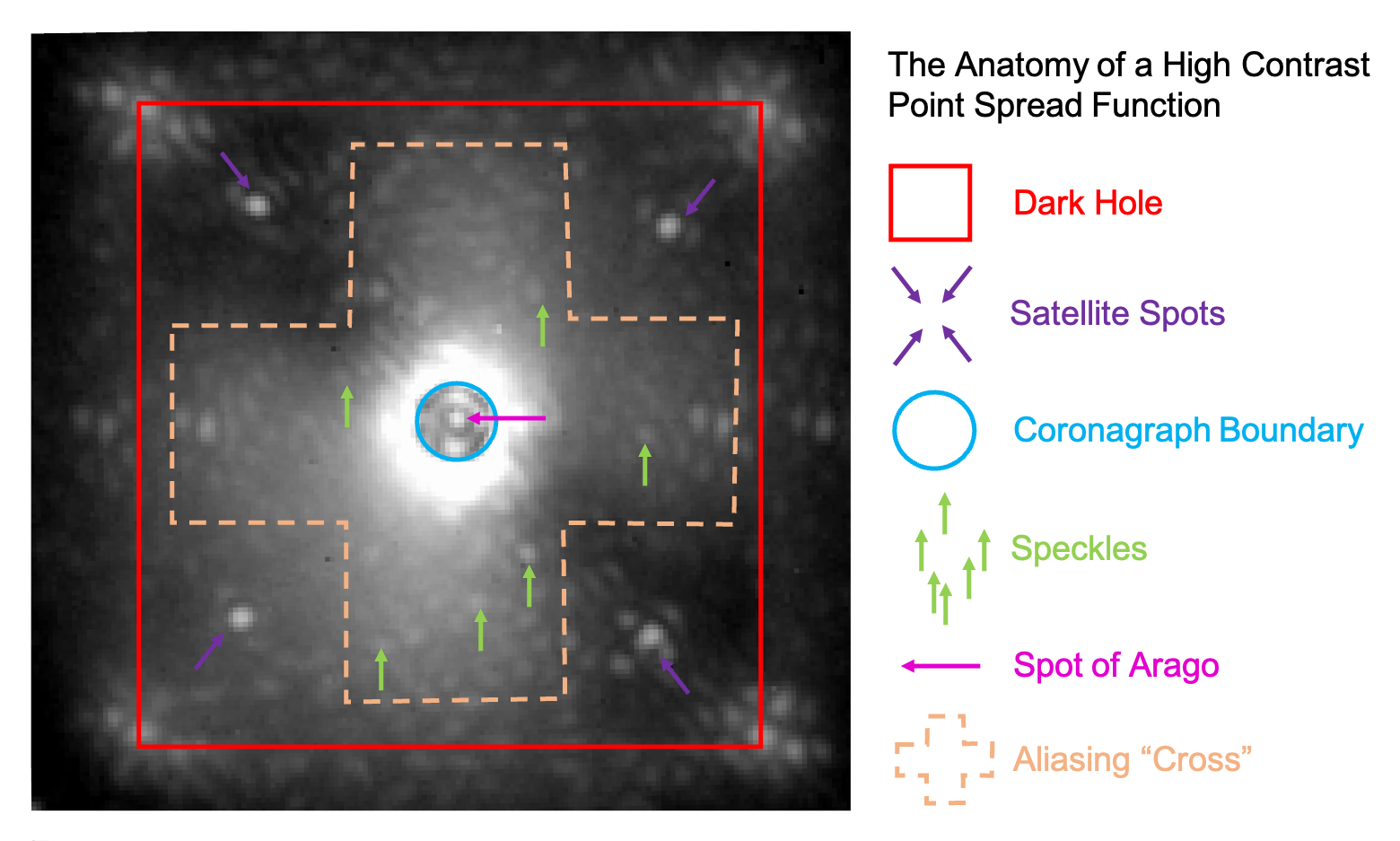}
    \caption{A raw high-contrast image from the Gemini Planet Imager, with various features labeled. GPI's square-shaped ``dark hole" (region of AO correction" is marked in red. Satellite Spots injected intentionally into the images by the apodizer are shown with purple arrows, and serve as photometric and astrometric references. The central star is obscured by the coronagraph, the edge of which is depicted in blue. Diffraction does introduce some light to the region ``underneath" the coronagraphic mask, including the ``Spot of Arago" at the center of the image, marked in magenta. Examples of speckles, which are distributed throughout the image but are concentrated near the edge of the coronagraphic mask, are marked in green. Individual high-contrast imagers have various unique features, such as GPI's ``aliasing cross" \rev{\citep[an optical effect caused by undersampling, see][]{Poyneer2016}}.}
    \label{fig:hcianatomy}
\end{figure}

\section{The Anatomy of a High Contrast Image \label{sec:HCIPSF}}

Unlike many other fields of astronomy, raw HCI images rarely contain any readily apparent raw signal from the target sources, even under aggressive hardware suppression of the stellar PSF. Post-processing is generally required to achieve the required contrast, and is covered in detail in Section \ref{sec:algos}. 
Nevertheless, the anatomy of a raw high-contrast image is important to understand in order to develop intuition for the range of artifacts that might survive into post-processing so they can be recognized and rejected as non-disk or non-planet signals. This section lays out the anatomy of a ``typical" coronagraphic high-contrast PSF, beginning with features at the center of the image and moving outward. 

\paragraph{Coronagraph and Spot of Arago --} First, the presence of a coronagraph in the beam results in a relative dearth of light at the center of the image. Generally the size of the coronagraphic mask can be discerned in raw images by the ring of bright diffracted starlight just beyond the outer edge of the coronagraph. Inside of this ring, the image is markedly darker, but there is often a single brighter spot at the center, the so-called ``spot of Arago" or ``Poisson spot", an artifact of Fresnel diffraction (and occasionally also airy rings surrounding it). This spot is not sufficiently bright to be used as a photometric or astrometric point of reference, however its detection and interpretation was central to our understanding of light as a wave and it thus has a very important role in the history of optics. 

\paragraph{Optical Aberrations --} The evolving atmosphere and the many optical elements of a high-contrast imaging instrument inevitably induce deviations in the PSF from the theoretical Airy Pattern of a circular aperture. Many of these aberrations can be sensed and corrected by the Adaptive Optics system, but imperfectly, such that some will survive into the final PSF, causing its shape to deviate from an Airy pattern and from image to image. 

\paragraph{Speckles --} The residual, uncorrected starlight that dominates raw high-contrast images generally comes in two forms. First, atmospheric or instrumental aberrations undetected or not fully corrected by the adaptive optics system manifest as ``speckles" (images, often aberrated, of the central star) at a range of locations in the PSF, but concentrated toward the optical axis. These evolve with the rapidly changing atmosphere, and blend into a diffuse halo of uncorrected starlight in most raw images (the so-called ''seeing halo"). For very short exposures, such speckles can be individually distinguished more readily, but in such cases they evolve quickly among images and thus rarely masquerade as planets in final PSF subtracted images. So-called ``quasi-static" speckles are likely created by optical aberrations in the instrument and evolve much more slowly, thus appear stably across multiple images and are more problematic. Various forms of active control are being developed to remove these quasi-static speckles \citep[e.g. ``speckle nulling",][]{Borde2006,Martinache2014} and many differential imaging processing techniques are designed specifically to distinguish quasi-static speckles from planets (see Section \ref{sec:diffim}).

\paragraph{Dark Hole/Control Region --} AO-corrected images also exhibit a boundary between the region of sensed wavefront aberration/AO correction and an uncorrected/unsensed region. This boundary definines the so-called ``dark hole" or ``control radius" of an AO system. The location of this boundary in the image plane is a direct consequence of the wavefront sensor's inability to perfectly sense all pupil plane wavefront aberrations. For example, there is a minimum size of wavefront aberrations that an AO system can detect and correct, set by the spacing of actuators, wavefront sensor optical component spacings (e.g. Shack Hartmann WFS), and/or wavefront sensing camera pixel scales (e.g. for a Pyramid WFS). Any spatial frequency smaller than this limit cannot be corrected by the AO system, and this pupil plane limit maps to a particular location in the image plane. Thus, the image reverts to seeing limited outside of the boundary of the dark-hole, resulting generally in an increase in the intensity of the seeing halo at its boundary. 

\paragraph{Wind Artifacts --} Wind, particularly high altitude wind, drastically affects the speed at which the incoming wavefront changes in time. AO systems therefore have a harder time 'keeping up' with aberrations along one axis of the PSF (the wind direction) than others, and the AO correction is therefore poorer along this axis.  In most modern HCI imagery, the wind direction can be inferred from an apparent elongation of the speckle pattern in the wind direction \rev{(i.e. there are more speckles in the halo along the wind direction, where the AO system is struggling to ``keep up")}. This additional uncorrected light introduces a difference in the achievable contrast in an image azimuthally, with planets/disks that align with wind artifacts more difficult to detect. 

\paragraph{Satellite Spots --} One practical consequence of coronagraphy is the loss of a direct measurement of the central star's astrometry and photometry. At the same time, photometric and astrometric characterization of substellar sources is dependent on these properties for the central star. For this reason, many modern HCI instruments inject reference ``satellite" spots into images at known locations and with known brightness ratios relative to the central star, either through a pupil plane optic custom-designed to inject them at certain locations and brightnesses or using manipulations of the deformable mirror of the telescope to produce them.  Once photometrically and astrometrically characterized \citep[e.g.][]{Wang2014}, these spots are sufficiently stable to allow them to serve as proxies for direct measurements of the location and brightness of the central star.

\paragraph{Instrument throughput}is a measure of the fraction of light entering the telescope aperture  at a certain wavelength that ultimately makes it onto the detector. It is determined in part by the number of reflecting and refracting elements in the optical path, each of which results in loss of a few percent of incoming light. The operating wavelength of the science camera and wavefront sensor is also a consideration. Generally wavefront sensors have operated at shorter, visible wavelengths and HCI cameras have operated in the NIR, enabling a dichroic to be used to separate these portions of the incoming light and minimize loss light at the science wavelength. The advent of Infrared wavefront sensors and visible light adaptive optics systems complicate this somewhat, to the extent that it can no longer be assumed generally that all light at the science wavelength is directed to the science camera, though clever combinations of filters and beamsplitters as well as usage of light that is otherwise discarded by the system (e.g. by the coronagraphic occulter) help to maximize throughput in these cases. 




\section{Differential Imaging Techniques \label{sec:diffim}}

\begin{figure}
    \centering
    \includegraphics[width=\textwidth]{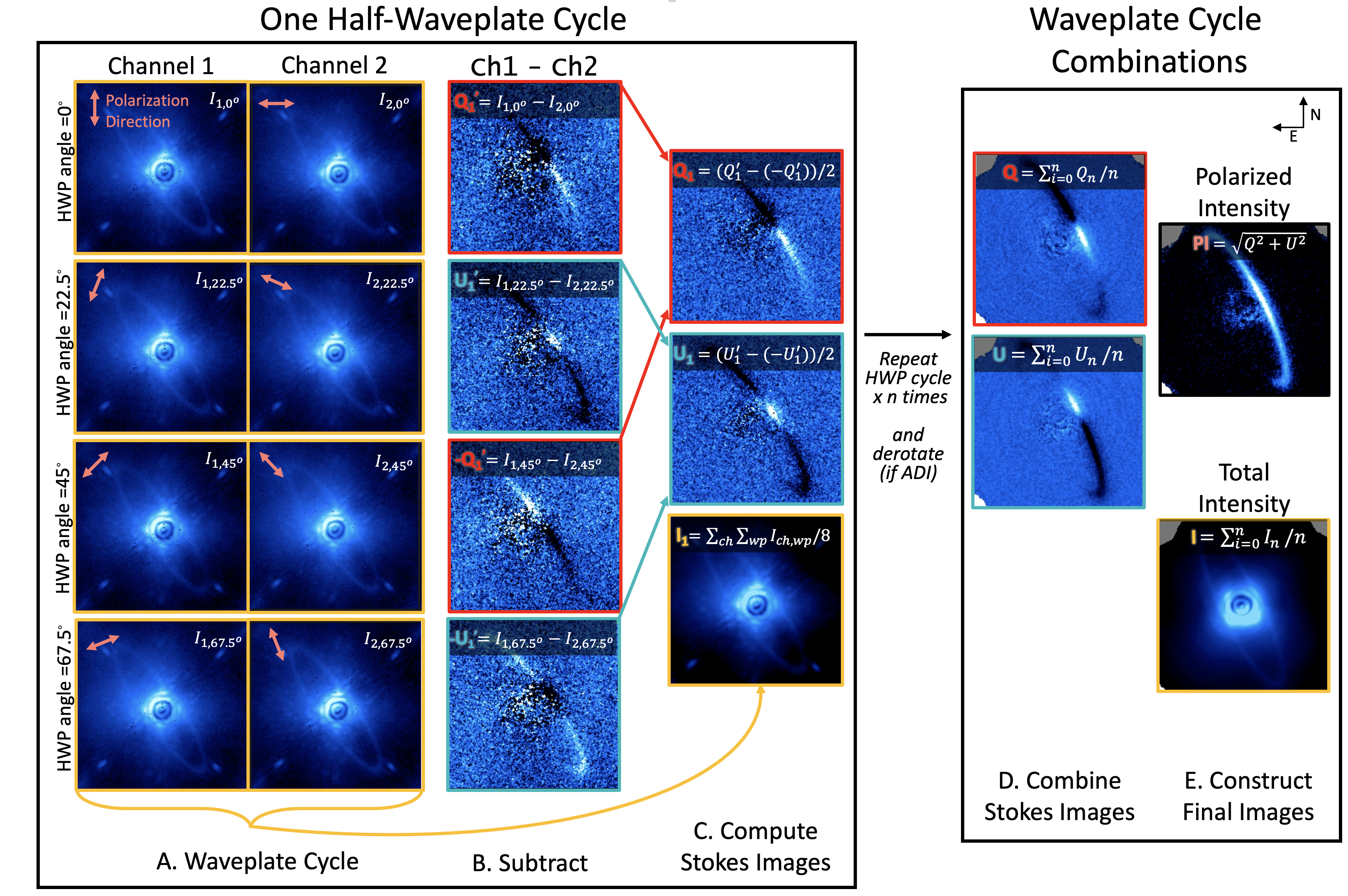}
    \caption{A schematic representation of the Polarized Differential Imaging (PDI) technique. Light from a disk-bearing star (in this case the debris disk host HR4796 A with the Gemini Planet Imager at K band) is split into two orthogonal polarization states (polarization vectors are indicated in coral in the figure), and these two ``Channels" (Column A's ``Channel 1" and ``Channel 2") are imaged simultaneously. A rotating Half-Waveplate (HWP) modulates the direction of both polarization directors by rotating 22.5 degrees between images, for a total of 4 pairs of polarized images, at orientations of 0, 22.5, 45, and 67.5$^{\circ}$. The two simultaneously-obtained orthogonal polarization channels are subtracted from one another (Column B). The subtractions for half-waveplate orientations 0 and 45$^{\circ}$ probe the Stokes Q parameter and its reverse.  The subtractions for half-waveplate orientations 22.5 and 67.5$^{\circ}$ probe the Stokes U parameter and its reverse, respectively. These independent probes of Stokes Q and U can be combined (Column C) to average over location specific artifacts. The dual channels of Column A can also be combined across all 4 waveplate orientations to yield a Stokes I (\revv{total intensity}) parameter image. This cycle of 4 waveplate orientations is repeated a number of times, often with Angular Differential Imaging (ADI) also employed (see Section \ref{sec:ADI}), allowing for individual Q and U images to be combined across a sequence (Column D). The square root of the sum of the squared Q and U images, is called the``Polarized Intensity" (PI) image (Column E). As can been seen in the figure, it easily isolates the (polarized) light of the disk from the (unpolarized) starlight, without the need for PSF subtraction. The combined total intensity image, on the other hand, is dominated by starlight.
    \label{fig:pdi}}
\end{figure}

Ultimately, even the best HCI hardware can only suppress starlight by 3 or 4 orders of magnitude in brightness, \rev{still 2-3 orders of magnitude too low in contrast relative to what is} required to image a hot young exo-Jupiter. Modern high-contrast imaging instruments rely on a number of clever data collection methodologies - collectively referred to as ``differential imaging" - to facilitate separation of starlight from planet/disk signal. When distilled to their essence, all differential imaging techniques are designed to leverage wavelengths, angular locations, other sources, or polarization states where companion light is faint or absent to estimate and subtract the PSF of the central star. These techniques are presented here in rough order of "aggressiveness" in estimating and removing the PSF of the central star. 

\subsection{Polarized Differential Imaging (PDI) \label{sec:PDI}}
Polarized Differential Imaging is the most common and successful technique for imaging circumstellar disk material in scattered light, and it is shown schematically in Figure \ref{fig:pdi}. It relies on the fact that light emitted directly from the central star is (generally) unpolarized. \rev{Dust grains in the circumstellar environment, on the other hand,} preferentially scatter starlight with a particular polarization geometry. \rev{Scattering is most efficient for} light with an electric field vector aligned orthogonal to both\rev{: (a)} the line of sight from the disk to earth and \rev{(b)}the vector connecting the dust grain and the central star. In principle, this means that a disk scattered light signal should dominate PDI images, and (unpolarized) stellar emission should be absent in polarized light images. 

PDI imaging leverages separation of incoming starlight \rev{according to the orientation of its electric field vector (i.e. it's linear polarization). An optic called a  Wollaston prism accomplishes this by passing incoming light through a material that has different indices of refraction for different linear polarization states. If a single Wollaston is used, the light is split into two beams with orthogonal polarizations (often called the ``ordinary" and ``extraordinary" beams), while a double Wollaston will yield four beams, adding redundancy that helps in removal of detector location-specific artifacts}.  The precise orientation of the orthogonal ordinary and extraordinary polarization vectors relative to the sky is manipulated to fully sample the polarized emission from the source by rotating an optic called a half- or quarter-wave plate, which modulates the orientation of the \rev{linear polarization state of incoming light} for the two channels.  This modulation (generally sequences of 4 angles - 0, 22.5, 45, 67.5 degrees) allows the images to be combined to yield the Stokes polarization vectors I, Q, and U \footnote{\rev{The Stokes vectors (a/k/a ``Stokes parameters") are a mathematical formalism used to describe the polarization state of light, namely: its total intensity (I), its linear polarization state (Q and U), and its circular polarization state (V). HCI instruments are not generally sensitive to the fourth Stokes vector V, so I will not discuss it here}}. Addition of images with orthogonal polarizations captures the \textit{unpolarized} intensity of the star, while subtractions yield either "Q" or "U" images, depending on the orientation of the waveplate. 
Q and U images are combined to isolate polarized light from the source via the equation $PI=\sqrt{Q^2+U^2}$. Each sequence of waveplate angles thus yields four images - I, Q, U, and PI. 

The angle of the polarization vector can also be extracted from these quantities as 
$$\theta_P=0.5\arctan\bigg(\frac{U}{Q}\bigg)$$
These vectors, when overplotted on images of a scattered light disk, demonstrate a characteristic centrosymmetric pattern. This is because of the preferred geometry of the scattering process where, as a reminder, the most efficient scattering occurs when a photon's electric field orientation ($\theta_p$) is orthogonal to both the line of sight and the vector connecting the scattering dust grain and star. 

Extraction of polarized signals is complicated somewhat by multiple scattering processes and the internal optics of the instrument. Internal reflections in the instrument result in  depolarization effects that vary with wavelength, incident angle, and the thickness and index of refraction of the optical components. This induces so-called ``instrumental polarization", which is typically estimated from observations of both unpolarized, disk-free stars and polarization standard stars. 

The simple picture of polarization presented above also assumes that each photon received was scattered by only a single small dust grain in the disk on its journey from star to disk to Earth. This is a reasonable assumption in many cases, but multiple scattering certainly occurs, and results in deviations in \rev{the centrosymmetry of polarization vectors, as well as differences in the characteristic pattern of positive and negative signal in Q and U images (often called a ``butterfly" pattern because the symmetric positive/negative lobes look a bit like butterfly wings)}. The inclination of the disk (i.e. whether emission is ``forward" or ``back" scattered) also impacts the efficiency of scattering, as do grain properties such as size, composition, and porosity. 

The most common variation on the process described above is to compute the so-called ``azimuthal" or ``local" Stokes Q and U vectors, often denoted $Q_{\phi}$ and $U_{\phi}$ \citep[e.g.][]{deBoer2020, Monnier2019} and defined as:

$$ Q_\phi = -\,Q \cos(2\phi) - U \sin(2\phi)$$
$$U_\phi = +\,Q \sin(2\phi) - U \cos(2\phi)$$

where $\phi$ is the azimuthal angle. This formulation has the advantage of concentrating signal with the expected polarization vector orientation into the $Q_{\phi}$ image, while the $U_{\phi}$ image becomes an estimate of the noise induced by multiple scattering and instrumental polarization.


\subsection{Reference Differential Imaging (RDI)}
The Reference Differential Imaging (RDI) technique utilizes images of stars other than the science target taken at other times to subtract starlight from a target image. It is an ideal approach when either (a) the PSF of a system is exceptionally stable, often the case for space-based observatories such as HST, or (b) the source being targeted has extended, symmetric features (e.g. a circumstellar disk) that might be subtracted by more aggressive algorithms that rely only on images of the target star \rev{of similar color\footnote{Similarity in color is important in RDI primarily because WFS and detector wavelength ranges are often different. Ideally, the reference star should be of similar \revv{\citep[or slightly higher,][]{Debes2019}} brightness at \textit{both} wavelengths so that its total flux on the detector (at the science wavelength) and the performance of the AO system (set by the star's brightness at the WFS wavelength) are similar.}} for reference (see next several sections).  Reference PSF libraries for RDI generally consist of images of many other stars taken at the target wavelength and in the same observing mode (e.g. same coronagraph) with the same instrument. In the case where a large library of reference images is available (e.g. a large HCI campaign, a well-established space telescope instrument), just a subset of the most highly correlated images may be chosen to construct a PSF. 

Some HCI observers, particularly of disks, regularly conduct PSF reference star observations as part of their efforts to observe a science target. PSF references are often chosen to be similar in location on the sky (so they can be observed interspersed with or immediately before or after the science target, at similar airmass), of similar apparent brightness at the wavelength of the WFS (so that the AO system performs similarly \footnote{One clever trick that some AO observers use is to ``pause" the AO control loop, slew the telescope to the reference star, and re-close it with all the same WFS algorithmic parameters in order to maximize this similarity}), and of similar color (so that the science image(s) have similar properties). \rev{Some modern HCI systems (SPHERE, MagAO-X) are equipped with ``star-hopping" modes that allow the AO loop to be paused on one target (e.g. the science target) and then re-closed once the telescope is pointed at another nearby target (e.g. the PSF reference star). This ensures maximal similarity in their PSFs.}  

\begin{figure}
    \centering
    \includegraphics[width=\textwidth]{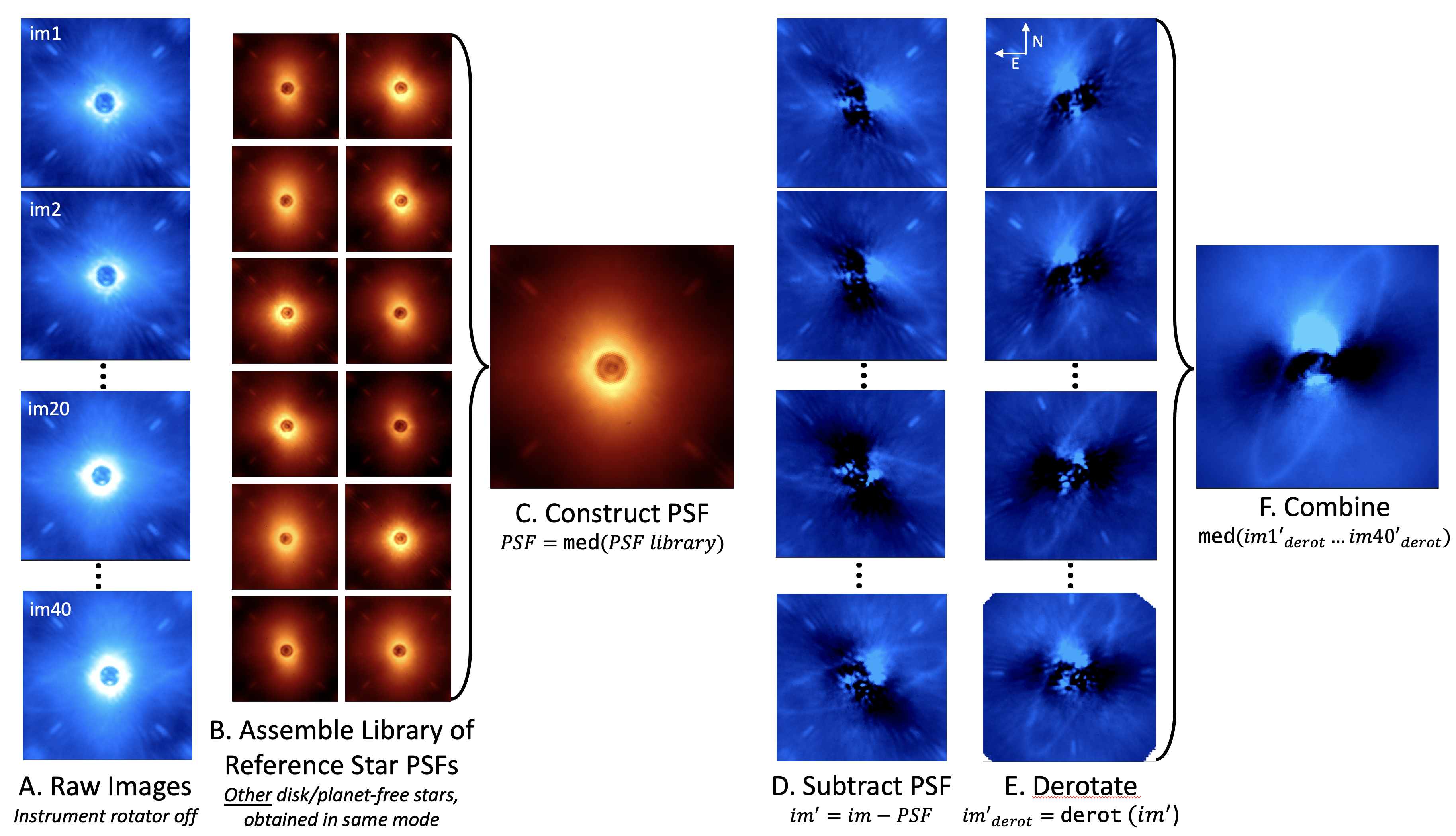}
    \caption{A schematic representation of the process of Reference Differential Imaging (RDI), in this case using Gemini Planet Imager H-band images of the debris-disk host HR4796A collapsed across all $\sim$40 wavelength channels of GPI. RDI utilizes a library of images of stars \textit{other than the science target} (Column B) obtained in the same observing mode. Generally, stars without any known disk or planet signal are chosen as references. These reference images can be combined simply (e.g. median combined, Column C) or used to build a custom PSF for each target image in the sequence (see Section \ref{sec:algos}). This PSF estimate is subtracted (Column D) to remove starlight in the image. In the case where the images were obtained with the instrument rotator off (typical for ground-based observing, see Section \ref{sec:ADI}), these subtracted images are rotated to a common on-sky orientation (Column E) and combined (Column F).}
    \label{fig:RDI}
\end{figure}

\subsection{Angular Differential Imaging (ADI) \label{sec:ADI}}
The Angular Differential Imaging (ADI) technique builds on the legacy of ``roll-subtraction" pioneered with the Hubble Space Telescope \citep[HST, e.g.][]{Schneider2014}. It leverages angular diversity to separate stable and quasi-stable PSF artifacts from true on-sky emission. It is predicated on the assumption that the instrumental PSF remains (relatively) stable in the frame of reference of the instrument throughout the image sequence, while true on-sky signal rotates with the sky. This allows the time series of PSF reference images to be leveraged for pattern matching or statistical combination to estimate the stellar PSF and remove it from each image. In practical terms, the quality of any ADI-based subtraction is generally a strong function of the amount of on-sky rotation of the source. For this reason, most direct imaging target observations are roughly centered around the time of that object's transit across the meridian, as this maximizes the amount of rotation achieved for a given amount of observing time. \rev{Rotation is essential to reduce a phenomenon called ``self-subtraction", in which the signal of a source (disk or planet) is present in a different but nearby location in the PSF image being subtracted, resulting in characteristic negative lobes on either side of the source where it has been subtracted from itself (hence the name).}

The simplest form of ADI, so-called ``classical" ADI (cADI), constructs a single PSF for subtraction from the median combination of all images in a time series, subtracting this median PSF from each image and then rotating these subtracted images to a common on-sky orientation. These PSF subtracted and re-oriented images are then combined, further suppressing the residual speckle field, which varies from image to image. 

\begin{figure}
    \centering
    \includegraphics[width=\textwidth]{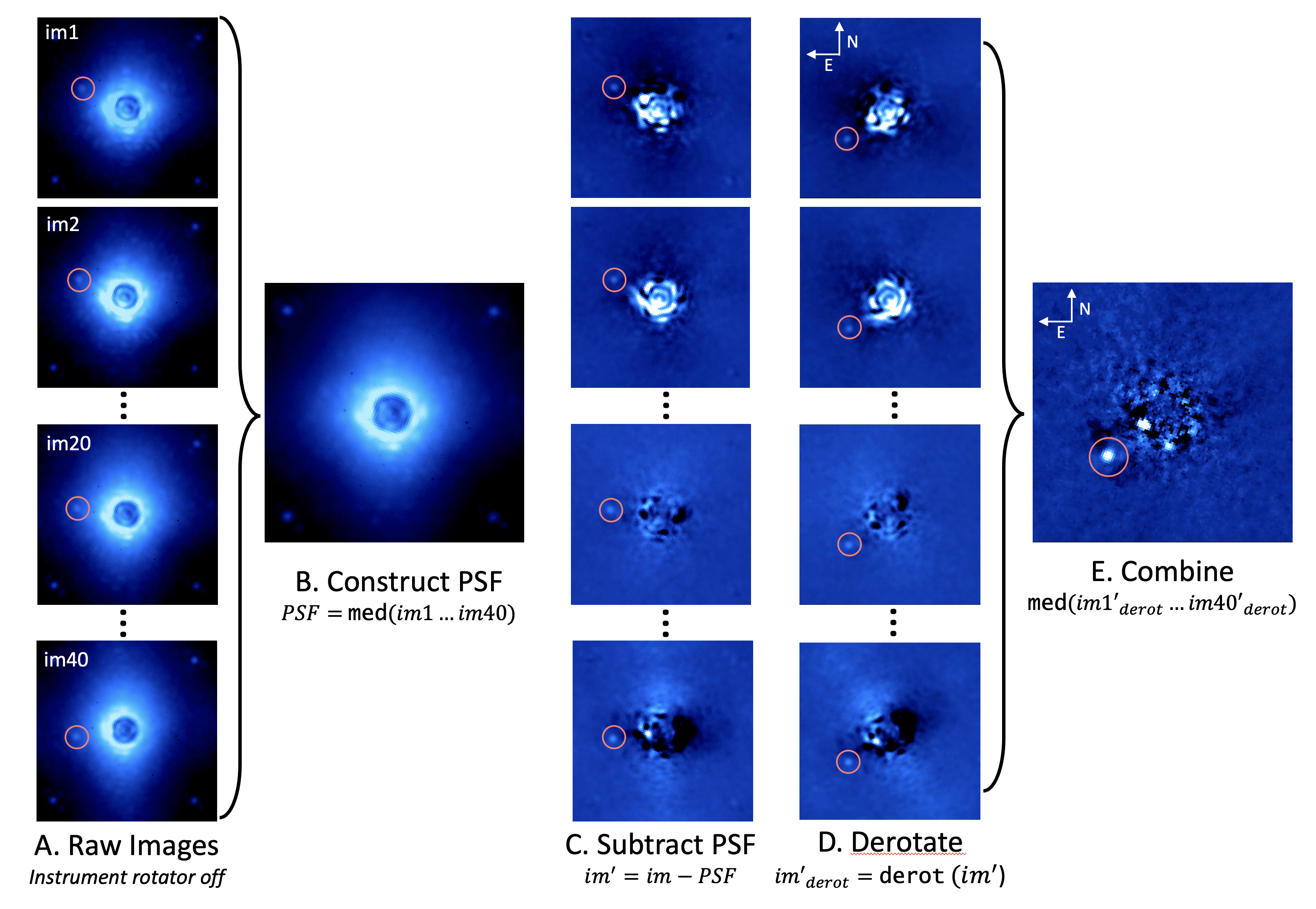}
    \caption{Illustration of the classical Angular Differential Imaging (cADI) technique using a sequence of 40  Gemini Planet Imager coronagraphic H-band (1.6$\mu$m) images of the planet host Beta Pictoris ($t_{exp}$=1min). Images (column A) are collected with the instrument rotator off, allowing the sky to rotate. The instrumental PSF (including any quasi-static speckles) remains relatively stable in the instrument frame, while real sources rotate with the sky relative to the instrument frame. The image sequence is median combined to create an instrumental PSF (column B), which is then subtracted from each image (column C), derotated to a common on sky orientation (column D), and median combined again (column E). In this case, the planet Beta Pictoris b (coral circle) is bright enough to be seen in individual exposures. It is not present in the PSF, as it rotates with the sky and is thus not in the same position throughout the image sequence. The median PSF is not a perfect PSF reference, and image-to-image variation can be seen in column C. However, derotating and median combining these imperfect subtracted images results in a very clear detection of the planet.}
    \label{fig:cadi}
\end{figure}

\subsection{Spectral Differential Imaging (SDI)}

HCI observing programs often aim not just to \textit{detect} exoplanets and circumstellar disks, but also to \textit{characterize} them, for which multiwavelength information is invaluable. Due to the many challenges of absolute photometric calibration in HCI (see Section \ref{sec:analysis}), characterization is best facilitated by obtaining simultaneous imagery at multiple wavelengths. Thus, many modern HCI instruments are so-called ``Integral Field Spectrographs" (IFSes). IFS instruments are used throughout Astronomy with a range of architectures, but in the case of HCI, they are generally of a fairly similar lenslet-based design. In lenslet-based IFSes, a grid of lenslets is placed in the focal/image plane of the optical system (not unlike the grid of lenslets placed in the puil plane of a SHWFS, see Section \ref{sec:wfs}), and then the lenslet spots are dispersed to produce a spectrum at each location. Each ``spectral pixel", or ``spaxel" (also referred to as a ``microspectrum"), contains spectral information \textit{at a particular location in the image plane}. Although it requires post-processing to do so, raw IFS images can be converted into a cube of resolved, multiwavelength images by using arclamps to connect locations along each microspectrum with specific wavelengths, fitting these locations photometrically by leveraging some knowledge of the instrumental PSF, and then placing that photometric value in an array at the appropriate spatial location relative to other values. A raw IFS HCI of the planet-host Beta Pictoris is shown in Figure \ref{fig:IFS} and broken down schematically. 

Spectral Differential Imaging takes advantage of differences in the spectral properties of planet and starlight. In particular, it leverages images at wavelengths where planets are generally dim (e.g. for methane dominated planetary atmospheres, at 1.5 and 1.7$\mu$m) to construct a PSF model that is largely uncontaminated by planet light, limiting self-subtraction. Because images are collected at multiple wavelengths contemporaneously, this circumvents some of the effects of a temporally-varying PSF. As a result, the library of reference images are often better matched to the target PSF.  

Most high-contrast SDI imaging to date has been done with an Integral Field Spectrograph such as GPI or SPHERE. These instruments separate the focal plane into a grid of so-called "spaxels" by focusing light on a grid of lenslets then passing the separated lenslet spots through a wavelength dispersing element to achieve a grid of microspectra. A wavelength solution is derived for each microspectrum based on observations of an internal arc lamp in the system, which are generally taken close in time to science observations as the wavelength solutions can be highly dependent on the flexure of the instrument.  The microspectra are used to extract the brightness at a given wavelength for each spatial location (spaxel) and are then combined to create a cube of contemporaneously-obtained multiwavelength images. This is shown schematically in Figure \ref{fig:IFS}.

\begin{figure}
    \centering
    \includegraphics[width=\textwidth]{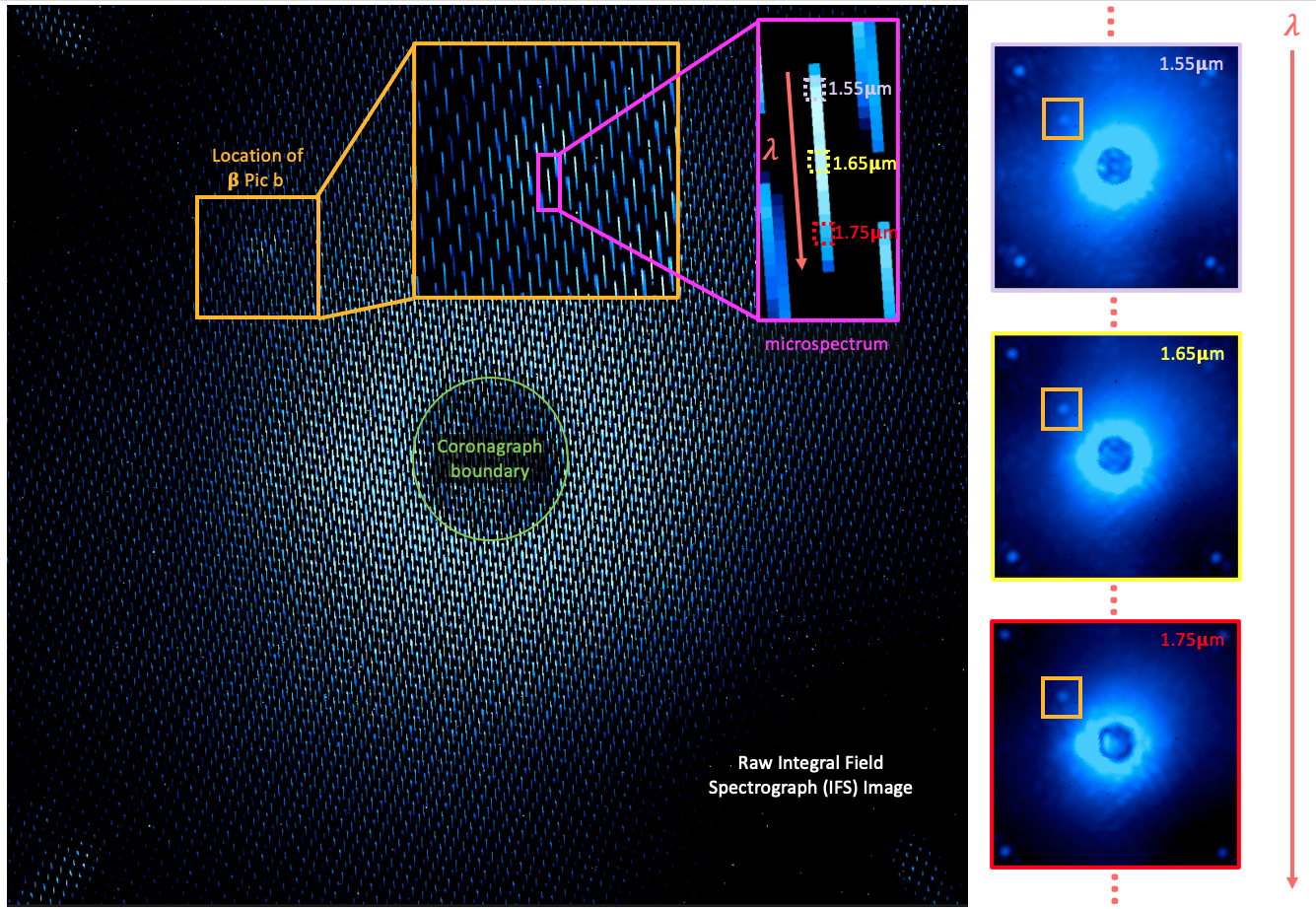}
    \caption{Schematic representation of the process of extracting a multiwavelength image cube from a single raw Integral Field Spectrograph (IFS) image. In this case, the background image is a raw H-band image of the star Beta Pictoris collected with the Gemini Planet Imager (GPI). Beta Pictoris has a known companion, Beta Pictoris b, whose light can be seen even in raw GPI images as a region of excess brightness in the wings of the stellar PSF, indicated in orange here. IFS instruments place a grid of lenslets in the focal plane, the light from each of which is passed through a dispersing element before reaching the detector. This creates an array of microspectra on the detector, one of which is highlighted in magenta here. Each microspectrum can be wavelength calibrated using arc lamps and its brightness extracted to create a single spectral pixel, or ``spaxel" for each wavelength (representative wavelengths of 1.55, 1.65, and 1.75$\mu$m indicated in cyan, yellow, and red on the microspectrum) and location in the image plane. These spaxels can be stitched together algorithmically to produce simultaneous images of the star at a number of wavelengths, creating a multiwavelength image cube rather than a single broadband image.}
    \label{fig:IFS}
\end{figure}

SDI can also  be implemented without an IFS by simply splitting incoming light into two beams with a 50/50 beamsplitter, dichroic, or Wollaston prism \footnote{A 50/50 beamsplitter splits light equally across a wide wavelength range. A dichroic is transmissive for some wavelengths and reflective for others, resulting in preservation of all of the intensity at a given wavelength. A Wollaston prism is similar to a 50/50 beamsplitter for the case of unpolarized input light -- it does not split light by wavelength, but rather by polarization state.}, and passing each beam through a different narrowband filter. This is sometimes called \textit{Simultaneous} Differential Imaging (still SDI). The SDI filter pairs lie  on- and off- of a spectral line of interest, and the most common lines used in today's high-contrast imaging campaigns are on- and off-methane in the NIR and on- and off- H$\alpha$ in the optical. In the case of young moving group stars (ages 10-300Myr), it is expected that planets will be faint or undetectable in the methane band due to absorption in giant planet atmospheres likely dominated by this gas, and brighter outside of the methane bands (see Figure \ref{fig:SDI}). H$\alpha$ differential imaging, on the other hand, leverages the fact that many younger (generally $<$10Myr) systems show evidence of ongoing accretion onto their central stars. The accreting material originates from and is processed through the circumstellar disk, meaning that any planets embedded in that disk are also likely to be actively accreting. One principal escape route for the energy of infalling material is radiation in hydrogen emission lines, particularly H$\alpha$, and we expect accreting protoplanets to be bright at this wavelength and faint or undetectable in the nearby continuum.

In terms of its utility as a tool to separate star and planet light, in its most generic form (what we might term ``classical" SDI imaging, shown schematically in Figure \ref{fig:SDI}) simply leverages the fact that the physical size of a stellar PSF on a detector is a function of wavelength. For simultaneously-acquired imagery at multiple wavelengths \rev{(i.e. A 3D cube of images with 2 spatial coordinates and 1 wavelength coordinate)}, this manifests as a magnifying effect as wavelength increases, and means that PSF features shift radially outward in detector coordinates, while true on-sky objects remain at the same position regardless of wavelength. Much like ADI angular rotation, the size of this effect is well-known (having a $\lambda$/D dependence), therefore it can be compensated for in post-processing. By rescaling (expanding shorter wavelength images or compressing longer wavelength ones) simultaneously-obtained images at multiple wavelengths to a common PSF scale, the wavelength-independent features of the PSF can be estimated. This rescaling alters the position of real objects in the images so that they are no longer in precisely the same location at all wavelengths, thus the rescaled images can be combined (e.g. via median or weighted-mean combination) to construct a relatively \footnote{I say relatively only because the difference in the position of a planet in wavelength-rescaled images is generally small compared to typical angular rotations for ADI processing, and more planet light is likely to survive into any estimated PSF.} planet-free PSF reference. This reference can be subtracted from the rescaled images and then the rescaling can be reversed to restore true on-sky coordinates, effectively realigning the planetary signals across wavelengths. These images can be collapsed in wavelength space to provide a robust planetary signal, enabling detection or astrometric characterization. More commonly, however, wavelengths are kept separate and combined across a sequence of multiple IFS images. This enables extraction of planet photometry at each wavelength to create a coarse spectrum, with a spectral resolution controlled by how many spectral channels can be extracted across the wavelength range of the IFS, generally a few dozen over a $\lessapprox$0.5$\mu$m wavelength range, for resolutions on the order of $\sim$25-100.

SDI processing is rarely used in isolation, and is rarely executed in the simple ``classical" sense described above. Instead, it almost invariably applies more sophisticated PSF estimation techniques to create custom PSFs for each image and wavelength within the image cube \rev{(i.e. using KLIP or another algorithm). Combination of SDI and ADI processing allows the user to leverage both angular and spectral diversity to identify reference images where planets might reasonably be expected to have moved enough to prevent their surviving into any combination (either through angular rotation or image rescaling).}

In addition to taking advantage of the physical rescaling of the instrumental PSF to identify images taken at the same or similar times to use as references, SDI processing also often involves the application of one or more planetary spectral templates to expand the reference library. For example, if we expect a planet with a methane-dominated atmosphere, such as the planet 51 Eridani b, then there are certain H-band wavelengths where we might expect methane absorption to make fainter planetary signals undetectable. We might leverage such wavelengths then as references, regardless of their wavelength separation from the image for which we are constructing a PSF.

\begin{figure}
    \centering
    \includegraphics[width=\textwidth]{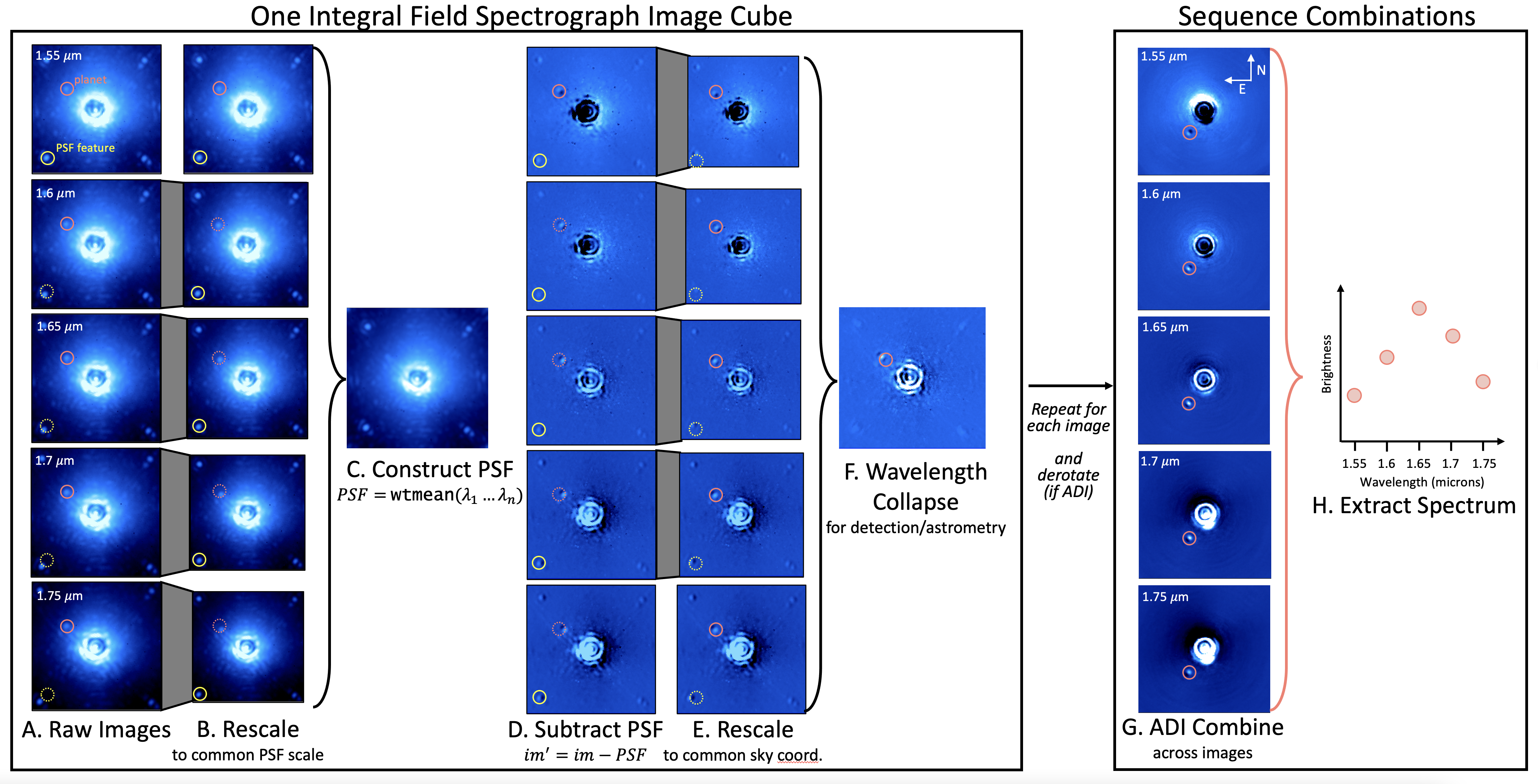}
    \caption{A schematic representation of the process of ``classical" Spectral Differential Imaging (SDI). Simultaneous images of a star are obtained at a range of wavelengths at once, in this case IFS images of the star Beta Pictoris obtained with the Gemini Planet Imager at H-band (1.5--1.75$\mu$m). A representative set of 5 of 37 total wavelengths \rev{from the 3D image cube (2 spatial, 1 wavelength dimension)} are shown in Column A, spanning a majority of the wavelength range. Each image is rescaled to compensate for the magnification of the stellar PSF with wavelength (Column B), placing instrumental PSF features on the same spatial scale (such as the satellite spots, one of which is indicated in yellow throughout). This rescaling, however, shifts the position of any real on-sky signal (such as the light from the planetary companion Beta Pic b, indicated in pink throughout). Rescaled images can then be combined (Column C) to create a relatively planet-free PSF (in this case by taking the weighted mean of the first and last few images in the rescaled image cube, where the planet light is farthest apart), which can be subtracted from each rescaled image (Column D) to remove a majority of the stellar signal. The rescaling must then be repeated in reverse (Column E) to re-align true on-sky signals before combination. Images can be combined in wavelength space to achieve detections or astrometric measurements (Column F), or the separate wavelengths can be retained and combined across a sequence of IFS images (Column G). Photometry of the planet can then be extracted from each combined image to construct a spectrum (Column H).}
    \label{fig:SDI}
\end{figure}

The size of a stellar PSF is a function of wavelength; it increases as the wavelength does. Raw SDI image cubes are therefore not initially good references for one another. Their spatial scales must first be adjusted to a common magnification in order to construct a PSF library. While this makes the instrumental PSFs of the multiwavelength images match, an effect of this rescaling is that the true on-sky spatial scale varies across the wavelength dimension of the reference images. This can result in \textbf{radial} self-subtraction of the planetary PSF when planet light at another (rescaled) wavelength makes it into the library of reference images. 

A distinct advantage of SDI is the acquisition of spectral information, which allows for atmospheric characterization \rev{of directly-imaged companions and composition analyses of circumstellar disks}. Although the mechanics of the technique are somewhat different and outside of the scope of this tutorial (relying on the placement of optical fibers on and off of the known location of a directly imaged companion), it's worth noting that medium- and high-resolution spectroscopy is increasingly being used to much more finely characterize the atmospheres of directly imaged companions. 

\begin{table}[]
\centering
\begin{tabular}{|L{16mm}|C{10mm}|L{21mm}|L{22mm}|L{45mm}|L{45mm}|}
\hline
\footnotesize
\textbf{Technique} & \textbf{Abbr.} & \textbf{Requirements} & \textbf{Best for} & \textbf{Advantages} & \textbf{Disadvantages} \\ \hline
Polarized Differential Imaging &
  PDI & Wollaston prism, rotating half waveplate &
  disk morphology and grain studies & - does not require PSF subtraction\newline - combined with total intensity imagery, probes disk grain properties & - instrumental polarization, multiple scattering effects difficult to isolate and remove\newline - forward/back-scattering can result in only one side of a disk being detectable \\ \hline
Reference Differential Imaging &
  RDI &
  reference star observations &
  detection and photometry\newline of extended disks &
  - allows for characterization of disks with arbitrary morphology, including face-on &
  - Difficult to achieve reference star observations with well-matched PSFs\newline - PSF star observations require additional observing time \\ \hline
Angular Differential Imaging &
  ADI &
  on-sky rotation &
  detection and photometry \newline of planets, narrow disk structures &
  - lots of on-sky rotation can enable more effective PSF subtraction close to star &
  - post-processed PSFs show azimuthal self-subtraction \\ \hline
Spectral Differential Imaging &
  SDI &
  spectrograph &
  spectral characterization\newline of planets, narrow disk structures &
- recovers spectral information, enabling characterization\newline - can leverage knowledge/assumptions of spectrum to improve PSF subtraction &
- post-processed PSFs show radial self-subtraction\newline - planet movement constraint range is narrower \\ \hline
\end{tabular}%
\caption{}
\label{tab:compare}
\end{table}

\section{Algorithms for High-Contrast Image Processing \label{sec:algos}}

 In addition to applying hardware (see Section \ref{sec:hardware}) techniques to suppress starlight and differential imaging (see Section \ref{sec:diffim}) techniques to facilitate separation of star and planet signal, most modern HCI efforts require additional post-processing beyond the ``classical" versions described in Section \ref{sec:diffim}, and the most common techniques to enable this are described in this section. 

\subsection{Filtering}

A common form of preprocessing for high contrast images is the application of so-called ``high-" or ``low-pass" filters to the data. This terminology refers to the spatial frequencies\footnote{This is a Fourier analysis term, and can be understood through the relation between pupil and image plane discussed previously. When an image undergoes Fourier transform, the intensity of the resulting 2D function can be related to the strength of various ``spatial frequencies" in the image. These can be thought of as maps of the degree of symmetry and typical size scale of variations in the intensity of the image.} that are the least suppressed by the application of the filtering algorithm - they ``pass through" the process relatively unscathed while other spatial frequencies are suppressed. A highpass filter allows through high spatial frequency signals such as narrow disk features and planets. A low-pass filter suppresses these signals while preserving extended structures such as the stellar halo or broad disk features. 

Highpass filters are applied to high-contrast imaging data before and/or after PSF subtraction. A simple example of a highpass filter is the so-called ``unsharp masking" technique, wherein an image is convolved with a simple kernel (often a gaussian), and then this smoothed image is subtracted from the original. High spatial frequency structures are drastically altered (spread across many more pixels than its original extent) by this convolution, while low spatial frequency structures remain largely unaltered. Thus, the subtraction suppresses these low-frequency signals while preserving the high-frequency structure. There are a range of additional algorithms/strategies used to achieve highpass filtering, many of which are applied to the Fourier transform of an image in the frequency domain, but all of which are designed to serve the same purpose.

\subsection{PSF Post-Processing}
A number of post-processing algorithms extend the concept of ``classical" differential imaging to construct \textit{custom} PSF models for \textit{every} image in a time series \textit{individually}, rather than adopting a single representative PSF for the entire image sequence. The two families of algorithm used most often are outlined below. Like ADI, RDI, and SDI, both rely on assembly of a library of reference images (often other images of the target itself taken in the same imaging sequence), and these images are used to construct the PSF model(s) for the target image. They rely on correlation between the target image and the other images in the reference library, weighting most heavily the images that are most closely correlated with the target image \footnote{\revv{In the case of RDI processing of a disk-hosting star, some portion of the image known to host disk signal may be excluded from consideration (masked) before computing these correlations. This ensures that regions of relatively pure stellar signal drive the choice of reference images for PSF model construction and minimizes oversubtraction of disk signal.}}. In this way, these algorithms are able to capture the time varying nature of the PSF and quasi-static speckles in images rather than relying on a single PSF for the entire image sequence. PSFs can be constructed for an entire image, or for azimuthally and/or radially divided subsections of the image, and these algorithms can be applied for ADI, SDI, RDI, and occasionally even PDI image processing. 

For these more advanced PSF-subtraction algorithms, restrictions are placed on which reference images are used to estimate the PSF for a given target image. The specific images in the sequence that are excluded and included in the reference library will change for each target image. Exclusion of images taken nearby in time or wavelength is done to limit the amount of planet light that survives into the PSF model. The consequence of planet signal appearing in the PSF models is azimuthal (ADI) and/or radial (SDI) self-subtraction. Their effects are shown in Figure \ref{fig:selfsub}. 

\begin{figure}
    \centering
    \includegraphics[width=\textwidth]{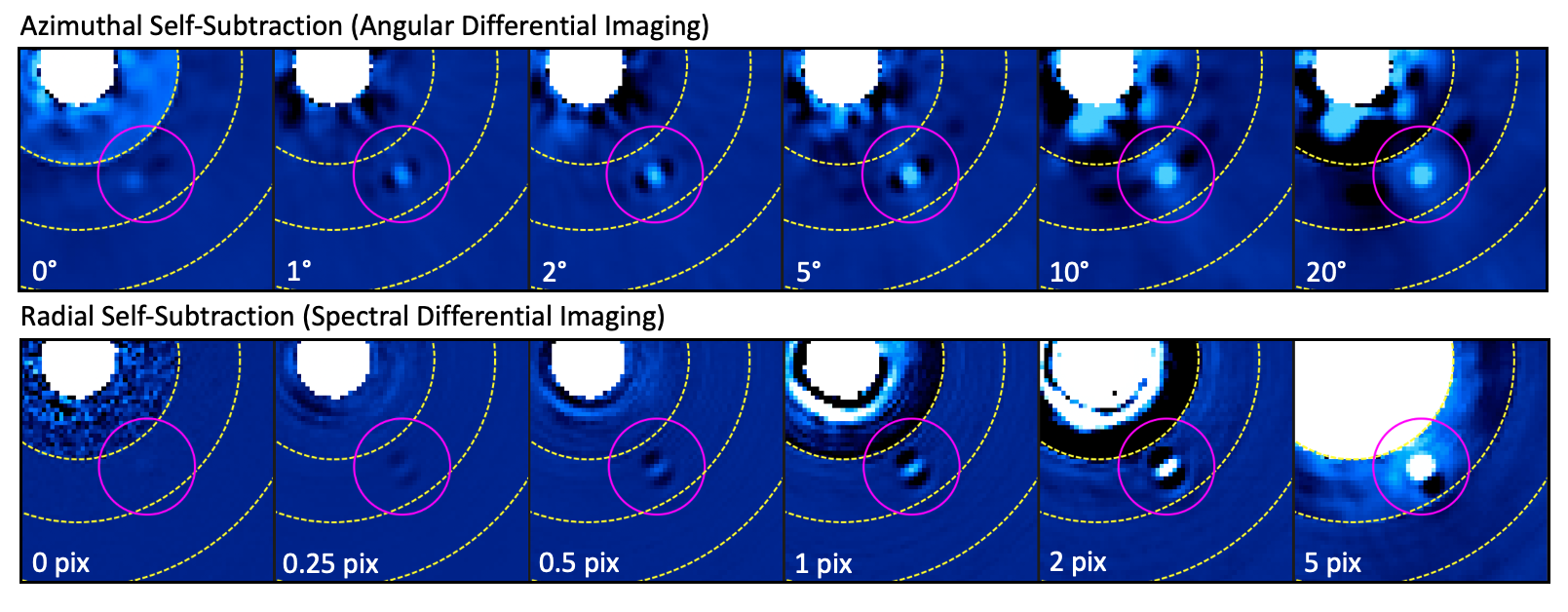}
    \caption{Azimuthal (top row) and radial (bottom row) self-subtraction of the planet Beta Pictoris b in KLIP-processed Gemini Planet Imager data. Azimuthal self-subtraction occurs in Angular Differential Imaging (ADI) when reference images where the planet's signal fully or partially overalaps its location in the target image are included in the PSF reference library. Radial self-subtratction occurs in Spectral Differential Imaging (SDI) when rescaled (to match the scale of the target image) PSFs at nearby wavelengths contain planet signal (shifted inward or outward in the rescaling) that overlaps that of the target image. KLIP includes a threshold for the amount of angular or physical motion that a planet at a given location must undergo (due to angular rotation for ADI and PSF rescaling for SDI) before another image in the sequence can be included in the reference library for PSF subtraction. This is a tunable parameter, and both top and bottom panels depict a sequence of very aggressive (no threshhold) to less aggressive reductions. An aggressive threshold generally provides better PSF subtraction (most evident at the center of the images) because the PSF library includes the images taken closest in time to the target image, but it also results in the highest degree of self-subtraction, evident in the characteristic dark-bright-dark of the post-processed planetary PSF, where the dark regions on either side of the core are referred to as ``self-subtraction" lobes. For the least aggressive reductions, self-subtraction is minimal (though the presence of the planet in the KL modes can be seen in the negative lobes extending azimuthally in the case of ADI (top row) and radially in the case of SDI (bottom row)), but PSF subtraction is also less effective. Fainter planets nearer the star may only be resolvable with more aggressive reductions.}
    \label{fig:selfsub}
\end{figure}

\subsubsection{KLIP} 
Karhunen Loeve Image Processing, or KLIP, is a statistical image processing technique in which images are converted to 1D column vectors and cross correlated with all other images in a time sequence. This application of Principal Component Analysis (PCA) allows for identification of common patterns (``principal components") in the image cube. 

PCA is used in a range of contexts inside and outside astronomy to reduce the dimensionality of data. A simple example of how it works is to imagine a 3D scatterplot with evident correlations among the x, y, and z axis quantities (as shown in Figure \ref{fig:PCA}). The x, y, and z coordinates are, in such a case, not particularly good descriptors of the overall data, in that it is only in combination that they can describe its variation. If we were to instead define a first principal component axis along the line of best fit, this single variable would capture the most distinct first order pattern in the data (it is the best single descriptor of the data's variance). If we were to add a second, perpendicular axis (in PCA each principal component is required to be orthogonal to all others), it would point in the direction of maximum scatter off the line of best fit, a good second order descriptor of the variance in the data.

\begin{figure}
    \centering
    \includegraphics[width=0.5\textwidth]{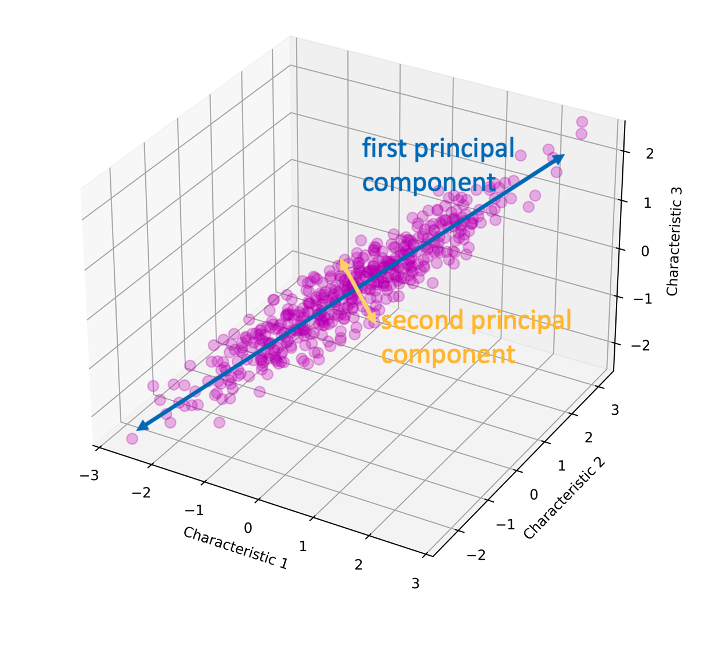}
    \caption{A simple visualization of Principal Component Analysis (PCA). To describe the position of any one data point in this dataset, one could specify three coordinates - it's x, y, and z location along the depicted axes. However, one could also provide a good (albeit imperfect) estimate of a point's location by simply providing a single coordinate - it's coordinate along a single vector that describes as much of the variation in the data as possible - the so-called ``first principal component" (depicted in blue here). If we also specified that point's location along an additional vector defined to be both: (a) orthogonal to the first principal component, and (b) pointing along the (orthogonal) direction describing the next greatest amount of variance in the data, this ``second principal component" (depicted here in yellow), together with the first, would provide an even better estimate of the point's location with only two coordinates. In high-contrast imaging, these patterns of covariance among images (principal components) can be used to model an image's Point Spread Function (PSF) \revv{using Karhunen-Loeve Image Processing (KLIP), which is a variant of Principal Component Analysis.}}
    \label{fig:PCA}
\end{figure}

It's difficult to extend this toy example conceptually into high numbers of dimensions, but the principal is the same - each additional orthogonal vector must be orthogonal to all others and is chosen to describe the maximum amount of additional variance in the data. Conceptually, in the case of PCA for HCI applications, this corresponds to patterns across many pixels that are present in the target image and that repeat frequently in the reference images. The first few principal components generally contain large scale PSF structures like core and halo, and the highest order principal components generally look like different realizations of the speckle pattern. Adding principal components to the model therefore increases its ``aggressiveness". This makes the likelihood of a well-matched PSF model higher, but also increases the likelihood that planet light will be oversubtracted or self-subtracted. 

\rev{KL modes are basically the principal components of a library of reference images that have been transformed into 1D arrays (albeit with some complexities that I will not cover in detail here). Once they} are computed, an individual image is ``projected" onto these KL modes, which in practice looks like a weighted linear combination of the principal components. KLIP algorithms lend themselves easily to returning models of varying complexity (different numbers of KL modes) simultaneously, so PSF subtractions can readily be generated with a range of aggressiveness and then compared. In such cases, low numbers of KL modes correspond to more conservative reductions, in that they (a) contain only the most widely varying PSF structures, and (b) result in relatively lower probability of any true circumstellar signals (disk, planet) being picked up as patterns that persist across images. The probability of these signals being picked up in the KL modes is much higher for spatially extended disks than for planetary point sources, so KLIP-ed disk images often use only a low number of KL modes (e.g. $<$10), while point-source reductions frequently use dozens to hundreds of KL modes. \rev{A schematic illustration of the KLIP process is shown in Figure \ref{fig:KLIP}}.

\begin{figure}
    \centering
    \includegraphics[width=\textwidth]{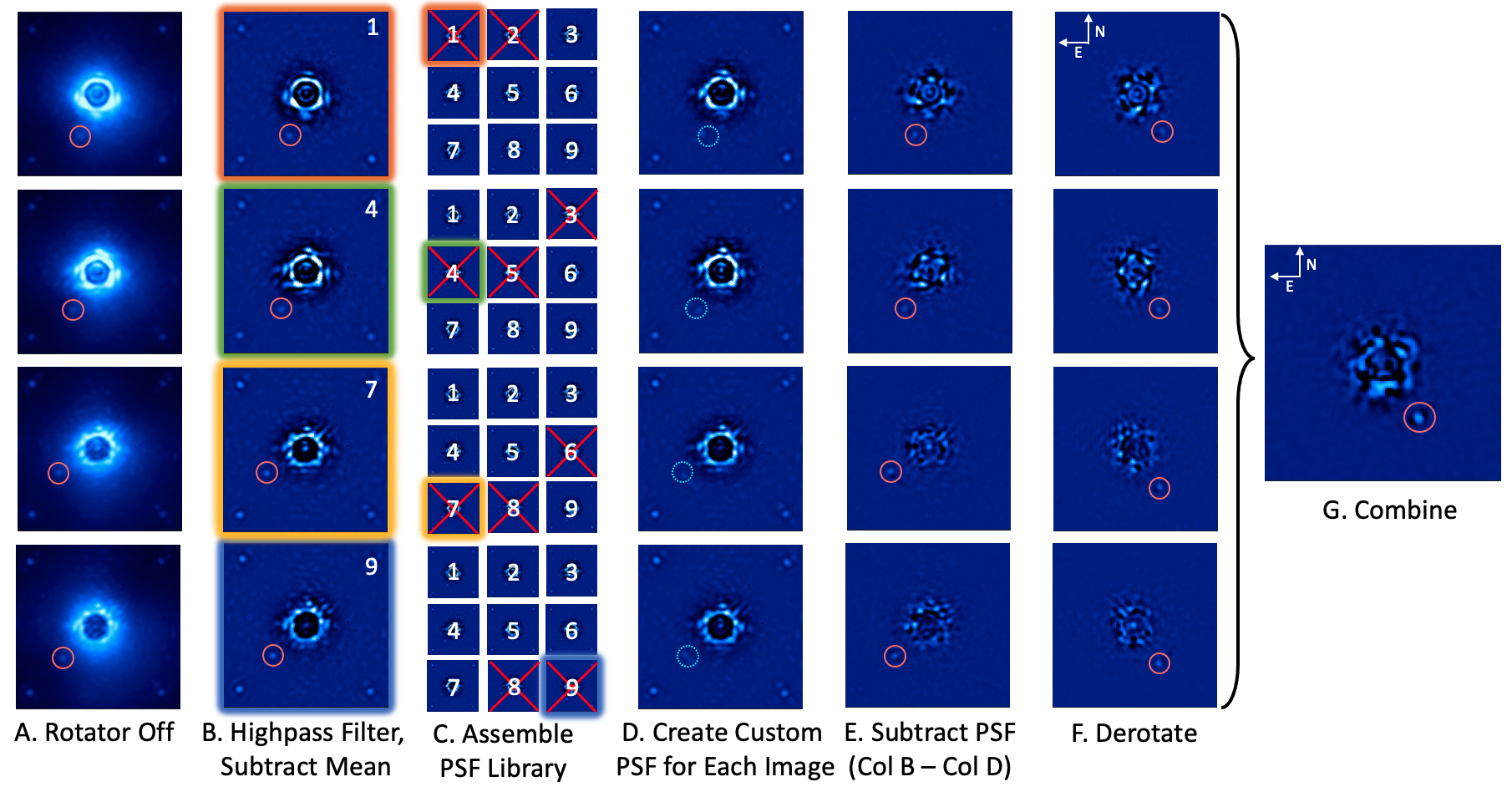}
    \caption{Illustration of the Karhounen-Loeve Image Processing (KLIP) technique. This technique can be applied to ADI, SDI, and RDI imagery, but is shown for the ADI case here. Like Figure \ref{fig:cadi}, this visualization utilizes a sequence of 40 Gemini Planet Imager coronagraphic H-band (1.6$\mu$m) images of the planet host Beta Pictoris. Images (column A) are collected with the instrument rotator off, allowing the sky to rotate. A collection of other images in the sequence (column B) are assembled for PSF modeling of \textbf{each} target image in the ADI sequence. Algorithmic controls determine the degree of ``aggressiveness" in including or excluding reference images taken near in time to the target image, where planetary signal may overlap (excluded images shown with red x symbols in column C). Principal Component Analysis of the reference library and target image allows for construction of one or more PSF models of tunable complexity (number of principal components in the model, column D depicts N=5 components). As in cADI, these models are subtracted from the target image (column E), derotated to a common on sky orientation (column F), and combined (column G) to reveal the planet.}
    \label{fig:KLIP}
\end{figure}

\subsubsection{LOCI}
The Locally Optimized Combinations of Images \citep[LOCI,][]{Lafreniere2007} technique constructs a PSF model by weighting and combining some number of images from the reference library as a PSF model for the target image. In its original form, the algorithm computes a least-squares fit to the target image using weighted linear combinations of the images in the reference library, with the goal of minimizing the residuals in the difference of the target image and the PSF model. Since it was originally developed, several enhancements have been made to the LOCI algorithm. A non-exhaustive list of these enhancements is provided below.
\paragraph{Template LOCI} \citep[TLOCI,][]{Marois2014}, was specifically designed for SDI imaging and its aim is to maximize the SNR of planets with a specified spectral shape. The user specifies a planet spectrum (e.g. flat, methane-dominated, etc.) and sets a threshold for the amount by which the planet's flux is allowed to be reduced by self-subtraction (due to both azimuthal FOV rotation with time and radial PSF magnification with wavelength). Using simulated planets, the amount of self-subtraction in each reference image is quantified. Images with predicted self-subtraction above a certain threshold are excluded from the reference library before the least-squares fit is computed.
\paragraph{Adaptive LOCI} \citep[ALOCI,][]{Currie2012} implements an additional step of subtracting the radial profile of the star (the seeing halo) so that the speckle patterns among images can be readily compared. It also constructs a reference library from only the most correlated reference images (those above a certain user-defined correlation threshold). \paragraph{The Signal to Noise Analysis Pipeline} \citep[SNAP,][]{Thompson2021} directly optimizes the non-linear signal-to-noise equation for a planet at a given location by dividing the vicinity of a planetary signal into an annular ``optimization region" and a smaller semi-annular ``subtraction region". Forward-modeled planet photometry, a vector of coefficients for the linear combination, and an estimate of the noise derived from those coefficients are optimized to maximize signal-to-noise ratio.  

\section{Comparison of Techniques}
Now that we've introduced both differential imaging techniques more generally and some of the processing algorithms that we use to extend them and isolate light from extremely faint circumstellar signals, we can compare the relative efficacy of and situations best suited to application of each technique. These considerations are summarized in Table \ref{tab:compare}. Another useful tool for comparing and contrasting techniques is examination of images generated with each technique for the same object. This is provided in Figure \ref{fig:compare} using both a very faint planetary signal \citep[that of 51 Eridani b,][]{Macintosh2015} and a debris disk whose narrowness facilitates recovery under all of the algorithms \citep[HR 4796A,][]{Arriaga2020}.

Both Table \ref{tab:compare} and Figure \ref{fig:compare} highlight the fact that choosing a technique requires consideration of many factors, including both the feasibility of the observations and the specific science aims. An important takeaway is that differential imaging techniques can be especially powerful in combination. For example, recovery of a disk signal in both PDI and RDI or ADI imaging allows for computation of the polarization fraction (P=PI/I), a sensitive probe of the disk's grain properties. For planets, recovery of signal via multiple processing techniques lends credence to its nature as a \textit{bona fide} planet. In other words, the various techniques neither compete with nor supersede one another - all are needed to construct a full picture.

\begin{figure}
    \centering
    \includegraphics[width=\textwidth]{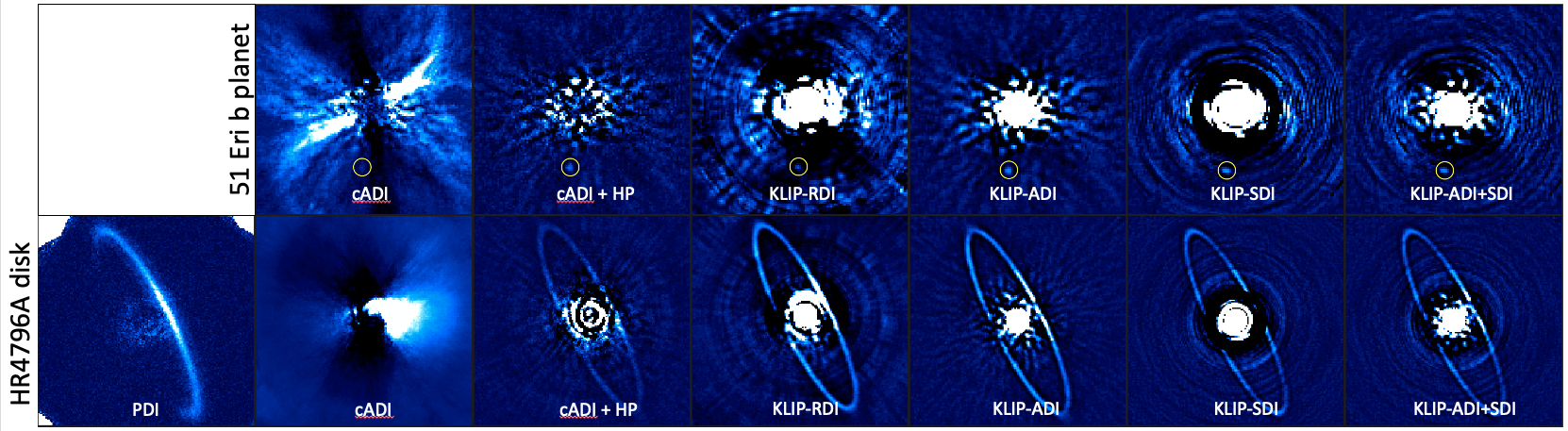}
    \caption{A young exoplanet (51~Eri~b, top row) and circumstellar disk (HR~4796~A, bottom row) reduced under a range of differential imaging techniques, from relatively conservative reductions at left to more aggressive reductions at right.}
    \label{fig:compare}
\end{figure}

\section{Analysis of High-Contrast Images \label{sec:analysis}}

\subsection{Contrast Measurement \label{sec:contrast}}
When reporting a high-contrast imaging \textbf{detection}, contrast is an important metric; however, it is also important in quantifying instrument performance in the case of a \textbf{non-detection}. Modern high-contrast imaging campaigns have surveyed a large number of young nearby stars with relatively few detections of exoplanets (e.g. 9$\substack{+5 \\ -4}$ planets for 5--13$M_{Jup}$ planets at separations of 10--100 au, \citealt{Nielsen2019}), though they have been more successful at detecting circumstellar disks \citep[polarized detection rates of $\sim$30-100\%, depending on selection criteria,][]{Esposito2020}. One of the main currencies of HCI surveys is therefore quantification of the instrumental performance, or limiting contrast, at a range of separations from each targeted star. This limiting contrast is a steep function of separation from the star, with lower achieved contrasts close to the star and higher contrasts at greater distances (see Figure \ref{fig:ccschem}). This means that a source at a given contrast is detectable in high-contrast images at a range of separations from the star, with bright sources being detectable at all but the tightest separations and the faint sources only detectable far from the star.

\begin{figure}
    \centering
    \includegraphics[width=0.75\textwidth]{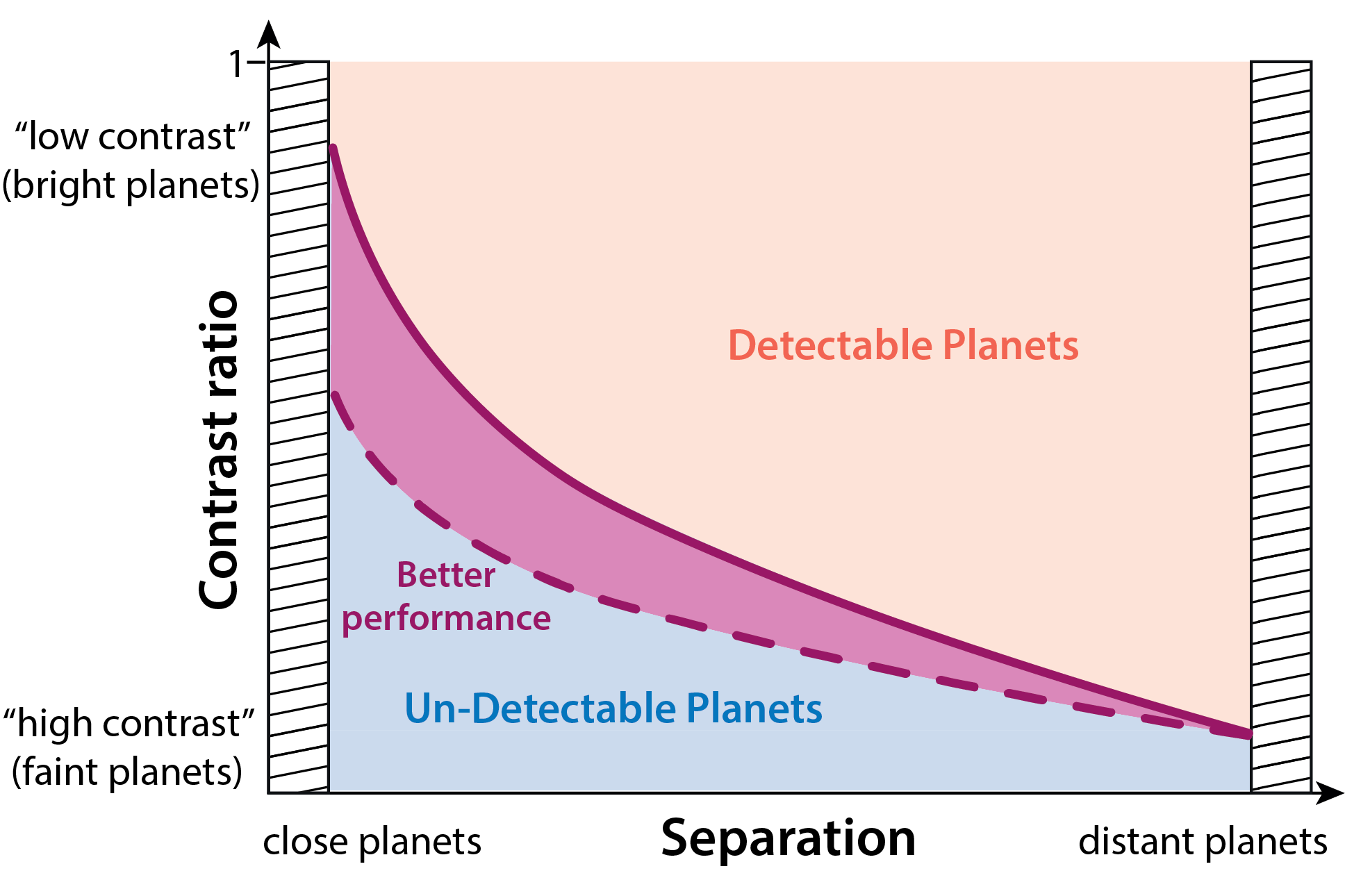}
    \caption{A schematic diagram illustrating how to read a contrast curve. At a given contrast and separation, a planet is detectable when it lies \textit{above} the curve. Achieved contrast is a steep function of separation from the central star, with only the brightest planets detectable at tight separations.}
    \label{fig:ccschem}
\end{figure}

These so-called ``contrast curves" therefore denote the detection \textbf{threshold} at each separation, with a few additional caveats and considerations. First and most importantly, many high-contrast imaging post-processing techniques (discussed in detail in section \ref{sec:algos}) do not conserve the flux of astronomical sources. This means that the ``raw" contrast, which is generally computed as 5 times the standard deviation of the noise at a given separation in the post-processed images, is not a true measure of the achieved sensitivity. 

In order to make a more accurate calculation, the algorithmic ``throughput" must be computed by injecting sources into the image at a range of separations and quantifying their recovered brightnesses. Throughput is defined as the ratio of an object's injected to recovered brightness (generally computed via the brightness of the peak pixel at the location of the source before and after PSF subtraction). Like contrast itself, it is a strong function of separation from the star. Throughput for most high-contrast imaging algorithms is low close to the star, meaning that source brightness is heavily suppressed in the PSF subtraction process, and approaches 1 at greater distances (meaning the planetary signal is relatively unaltered by PSF subtraction). The best estimate of recoverable planet brightness is therefore the 5$\sigma$ noise level of the image \textit{divided by} the instrument throughput at each separation from the star. This is sometimes called the ``throughput--corrected" contrast, but is most often just referred to as ``the contrast".

When computing throughput, an important consideration is overlap/crosstalk between injected sources, which can result in incorrect estimates. As sources can overlap both azimuthally and radially, the general approach for point source detection limits has been to inject false planets in an outwardly spiraling pattern with appropriate separations radially and azimuthally. Computation of throughput also requires a choice of injected contrast for each false source. Generally, a low to moderate contrast is chosen and set uniformly throughout the injected planet spiral so that recovery is assured, however it is likely that injected object throughput is, at least to some extent, a function of brightness.

\begin{figure}
    \begin{center}
    \includegraphics[width=0.8\textwidth]{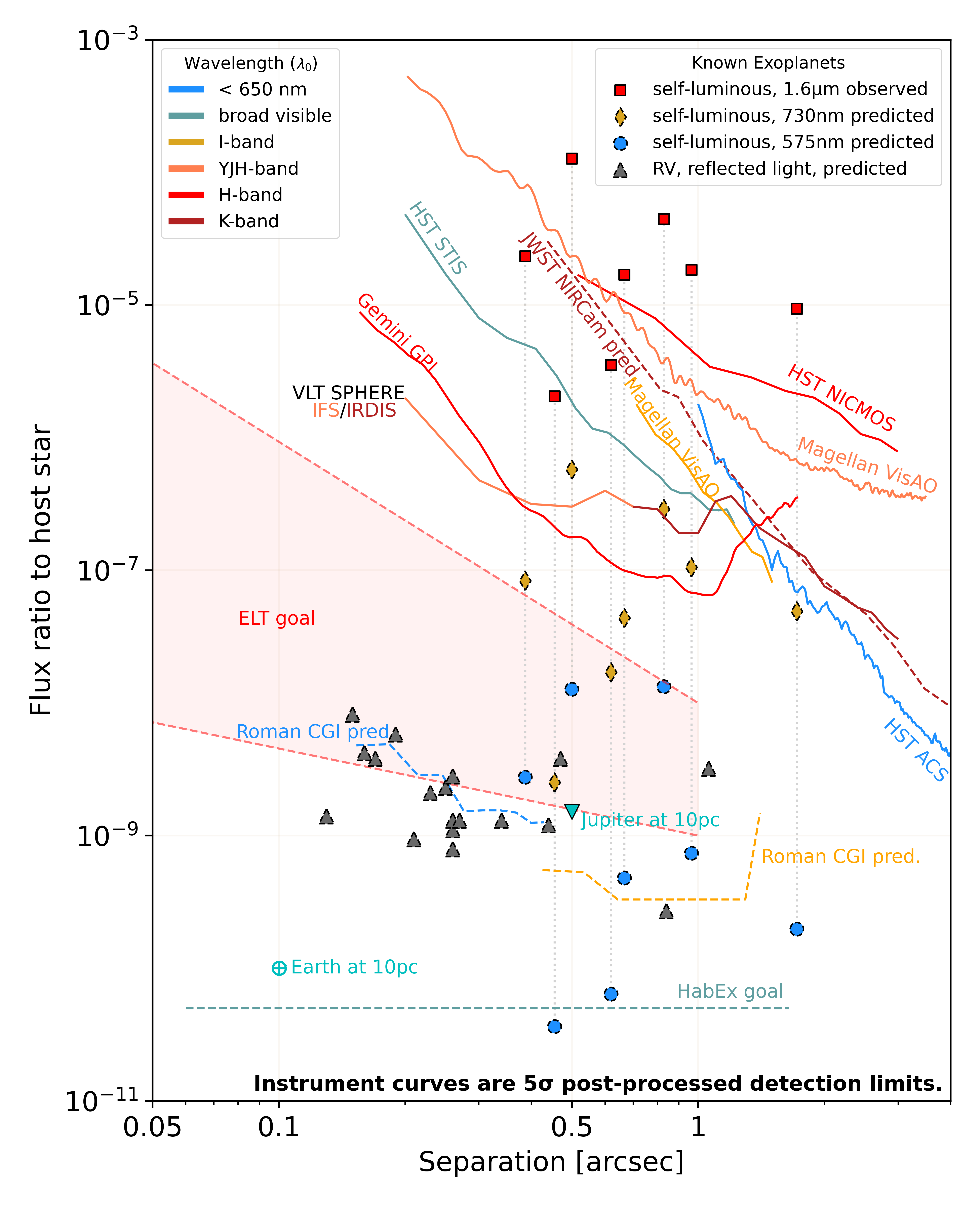}
    \caption{Demonstrated (solid lines) and predicted (dashed lines) contrast performance of various current and future HCI instruments. Lines and points are color coded by wavelength of observation. Points indicate both detected (solid outline) and simulated (dashed outline) planets. (\textit{code and data source: V. Bailey})\textsuperscript{a} \label{fig:cclandscape}}
    \end{center}
    \scriptsize{\textsuperscript{a} Alphabetical descriptions of plot elements. \textbf{ELT goal}: Possible range of near-IR post-processed detection limits for next generation extremely large telescopes. 
    \textbf{HabEx}: Goal 5$\sigma$ post-processed contrast.  IWA $\sim$ 2.5 $\lambda$/D @ 450nm; OWA $\sim$ 32 $\lambda$/D @ 1$\mu$m (source: B. Mennesson)
    \textbf{JWST NIRCAM}: simulated 5$\sigma$ post-processed [roll-subtraction] contrast curve for F210M-band. The model observation consists of 2x1hr rolls ($\pm$5$^o$), with a 10mas pointing uncertainty and a 10nm differential WFE. \rev{On sky JWST performance indicates that the true limit is lower still} (source: \citet{Beichman2010}).
    \textbf{HST NICMOS}: Best 5$\sigma$ post-processed [KLIP + match filter] contrast curve for F160W-band from the HST ALICE program (source: \citet{Choquet2014})
    \textbf{HST STIS}:  Bar5 coronagraph 5$\sigma$ post-processed [KLIP] contrast curve; 162sec exposure, bandpass $\sim$200-1030nm.(source: STIS handbook)
    \textbf{HST ACS}: 5$\sigma$ post-processed [simple image difference] contrast curve of 2x100sec Arcturus observation in F606W with 1.8" occulter. (source: J. Krist)
    \textbf{SPHERE}: 5 sigma post-processed [SDI] contrast curve for a $\sim$1hr integration on Sirius. At separations $<$0$\farcs$7 the curve is for IFS YJH, while $>$0$\farcs$7 is IRDIS K12 (Source: \citet{Vigan2015})
    \textbf{GPI}: 5$\sigma$ post-processed [KLIP + forward model match filter] contrast curve for H-band IFS mode, 1hr integration. Calculated from an 11min H-band IFS observation of Sirius. (source: B. Macintosh).
    \textbf{Roman CGI narrow FOV}: Modeled 5$\sigma$ post-processed [RDI, fpp=2] contrast curve for Band 1 imaging of a V=5 G0V star with the HLC coronagraph. Integration time is  10000hr (source: B. Nemati)
    \textbf{Roman CGI wide FOV}: Modeled 5$\sigma$ post-processed [RDI, fpp=2] contrast curve for Band 4 imaging of a V=5 G0V star with the SPC wide FOV coronagraph, based on OS9. Integration time is  10000hr (source: B. Nemati).
    \textbf{DI}: Selected self-luminous Directly Imaged (DI) exoplanets with known H-band contrasts. Predicted fluxes at Bands  1 and 3 are either from B. Lacy or from either COND (Teff$<$=1200K) or BT-SETTL models.
    \textbf{Earth, Jupiter}: simulated at quadrature as seen from 10 pc. (Jupiter albedo: 0.52 \citet{Traub2010})
    \textbf{RV}: All planets from NASA exoplanet archive with a semi-major axis of 0.12-1.4", mass $>$ 0.25 M$_{jup}$, and host star V mag$<$7. Lambertian flux ratio assumes: radius = 1 R$_{jup}$, geometric albedo = 0.5, circular orbit, inclination = 90.0, and angle of 0.0 degrees from the ascending node.}     

\end{figure}

\subsection{Limitations of Contrast as a Metric}

Contrast curves have several limitations. First, \rev{they are sensitive to post-processing choices (e.g. KLIP parameters), therefore optimization can be computationally intensive.} Second, they generally assume azimuthal symmetry in the sensitivity of post-processed images where in reality, stellar PSFs often have azimuthally dependent structure. One common example of this is the so-called ``wind butterfly" effect wherein lobes of higher noise/lower contrast are apparent on either side of a star in the direction of the wind in high-contrast images. This means that neither noise nor algorithmic throughput is truly azimuthally symmetric. One way to mitigate this is to \rev{inject false planetary signals at various locations azimuthally and measure how well they are recovered at each orientation}. For example, one might inject three spirals of false point sources with the spiral clocked by 120$^\circ$ each time in order to more fully sample the azimuthal variation in throughput. 

A further complication is in the definition of the ``noise" in post-processed images. The most standard metric is the standard deviation of the post-processed image computed in small concentric annuli extending outward from the star. The convention in high-contrast imaging is to consider sources whose peak recovered brightness is at least 5 times above the noise level to be robust detections, and objects in the 3-5$\sigma$ range to be marginal. Many contrast curves reported in the literature are so-called ``5$\sigma$" contrast curves, but 3 or even 1$\sigma$ curves are also sometimes reported and one must be careful to understand and correct for any differences when comparing contrasts among surveys. To put it plainly, all contrast curves should be interpreted as relatively rough and fuzzy boundaries between detectable and undetectable planets. 

One final consideration in computing and interpreting noise in a post-processed image is that the dominant noise source close to the star is stellar speckles. In this speckle-dominated regime, there is a strong correlation between flux in adjacent pixels, since the stellar PSF has a width of several to many pixels. This has led to a best practice of implementing t-distribution rather than Gaussian noise statistics at tight separations,  accounting for the small number of independent samples close to the star. In practice, this means dividing the computed standard deviation at a given separation by the factor $\sqrt{1+1/n_2}$ \citep{Mawet2014}, where $n_2$ is the number of independent noise realizations at that separation ($\sim2\pi r/FWHM$).

In summary, there are several important questions to ask oneself when studying a contrast curve.
\begin{enumerate}
    \item Is it throughput corrected? If not, remember that the true limit is likely at lower contrast (a higher curve).
    \item By what factor has the noise level been multiplied (1, 3, 5)? If less than five, recall that objects near the curve might be considered marginal or non-detections.
    \item Has the noise level been corrected to reflect appropriate noise statistics near the star? If not, the true limit may be a steeper function of separation from the star than depicted.
    \item How azimuthally symmetric is the post-processed image? If azimuthal structure is apparent, the curve should be interpreted as an average. In some parts of the image, objects below the curve may be detectable; in others, objects above the curve may be undetectable.
\end{enumerate}

Furthermore, one must keep in mind when comparing contrast curves between studies and instruments that these choices may not be uniform among them and the curves may not be directly comparable. For these reasons, it is important when planning observations and interpreting detections (or non-detections) relative to contrast performance, to carefully read contrast curve descriptions and discern these important details. You may practice contrast curve comparison and parsing of these details by perusing Figure \ref{fig:cclandscape}, which compares demonstrated and expected contrast of a range of current and future HCI instruments in several wavelength regimes. 

\subsubsection{Aside: Contrast Curves for Disk Detections}
Many of the points in the discussion above are altered or invalid for extended sources. Throughput, for example, is extremely difficult to compute for disks when their azimuthal and/or radial extent is large. Generally speaking, HCI disk detections utilize more conservative post-processing algorithms and observing techniques such as RDI for which throughput is much higher.

\subsection{Signal-to-Noise Calculation}
Signal-to-noise maps are standard in all fields of astronomy. In the case of direct imaging of point sources through PSF subtraction, there are several subtelties in computing them. First, the post-processed planetary PSF has characteristic ``self-subtraction lobes" on either side of the planetary core. These are caused by the presence of the planet at different azimuthal angles in the reference library, and therefore KL modes. The region containing the planetary core and self-subtraction lobes needs to be excluded in order to robustly estimate the noise at comparable radial separation from the star. This is typically done by masking this region and computing the standard deviation of the remaining pixels at a given radial separation. The nature of the speckle-dominated region of the PSF also means that independent samples of the noise at a given radial separation are defined by the size of a speckle (the PSF FWHM), leaving relatively few independent noise samples at tight radial separations and requiring t-distribution noise statistics \citep{Mawet2014}. 
One simple correction that is applied is to mask the region of planetary signal when computing the noise statistics. This results in a better estimate of the true noise level. 

\subsection{Astrometric, Photometric, and Spectral Extraction}
PSF-subtraction techniques, while powerful for isolating faint signals, complicate the extraction of accurate astrometry, photometry, and spectra from a detected object. At the most basic level, this is because the process of PSF subtraction does not conserve the original planet signal. 

A number of strategies are used to mitigate these complications and extract robust estimates of planetary photometric, astrometric, and spectral signals in HCI. 

\paragraph{False Planet Injection} Injection and recovery of false planet signals in the image helps to quantify the amount of planetary signal lost during image processing (as described in Section \ref{sec:contrast}). This is used in turn to correct photometry and estimate the true broadband intensity of planet light at a given wavelength. Similarly, false planets can be used to quantify astrometric and photometric uncertainties, often by injecting them into raw images at the same or similar radius as planet candidate(s) and utilizing the statistics of their recovered vs. injected locations and fluxes to quantify uncertainty on astrometry and photometry of the companion. 

\paragraph{Forward Modeling} Injection of a model companion or disk into raw images and examination of its morphology, astrometry, and photometry in post-processed images, is known as ``forward modeling". The properties of these false planets (brightness, location, fwhm) or disks (extent, inclination, radial brightness distribution) are iterated upon and the forward models compared to post-processed data. \rev{This process is essential in interpreting post-processed images, which suffer from both self-subtraction (see "Signal-to-Noise Calculation" above) and so-called ``over-subtraction", in which some of the planet or disk signal is flagged by the algorithm as noise and subtracted.}. Generally, forward models are tuned by attempting to minimize residuals in the difference of the PSF subtracted image and the forward modeled image. In many cases, models are injected not into the target image sequence, but into a reference image sequence or at a wavelength in the target sequence at which the signal is absent or minimized.  Post-processed signals are dependent on their azimuthal and radial location, and on the precise PSF, which is wavelength dependent, so neither of these techniques provides a perfect match. However, \citet{Pueyo2016} showed that a post-processed PSF can also be modeled for a particular location mathematically, without altering the original images, by propagating a perturbation to the covariance matrix forward through the algorithm (KLIP or LOCI). This removes the problem of mismatch by constructing a forward model at the same location and wavelength, and the authors demonstrated its ability to boost the accuracy of spectral extraction. Inferences made via forward-modeling are, however, limited by our ability to accurately model the true planet or disk signal, which is particularly difficult for complex off-axis or time-varying PSFs and non-axisymmetric disk structures. Nevertheless, post-processed PSFs, by virtue of our precise knowledge of their constructed photometry and astrometry, are powerful probes of the effects of PSF subtraction on the properties of real signals.

\paragraph{Negative Planets} Another robust technique for determining planetary flux and location is to inject \textbf{negative} false planets into the raw image sequence at the location of the planet candidate, effectively canceling its signal. The residuals following PSF-subtraction are then minimized to determine a best fit. Although this results in quite robust photometry and astrometry estimates, arguably better than using forward modeling, it is computationally intensive and uncertainties on this technique are harder to estimate. Often observers assign error bars ``by eye" to capture the range of values that result in good subtractions. For example, an appropriate flux scaling should result in near-zero residuals and not clear over- or under-subtractions (i.e. clear residual planetary excess or a clear residual negative signal at the planet location).

\section{Potential Sources of False Positives \label{sec:falsepos}}

Direct imaging detections are intrinsically difficult, testing the limits of current technology, and there are a range of both astrophysical and instrumental false positive possibilities. 

\subsection{Background Objects}

 One astrophysical false positive that can mimic a directly imaged companion signal is the coincidental alignment of a distant background source with a young star. This scenario is a possibility any time a faint point source is detected near a young star, thus it is among the first forms of vetting that all candidate planets are subjected to. For an initial single epoch detection, there are two important pieces of information that are used to assess the probability of a candidate being a background source - (1) the proximity of the target star to the galactic plane and (2) its spectrum. Coincidental alignments are much more common in the galactic plane, so the probability of false positives is higher in this case. As the most common background objects masquerading as planet candidates are distant red giants, spectral information - either true spectra or NIR colors - is also crucial in assessing the probability that a faint apparent companion is truly a young planet or brown dwarf. 
 
 With a few notable exceptions \citep[e.g. 51 Eri,][, an object whose methane-dominated spectum made its planetary nature clear from the outset]{Macintosh2015}, planet candidates are rarely announced until they have undergone an additional form of vetting - that of common proper motion with their host stars. Because the targets of direct imaging campaigns are close (generally $<$50pc), a necessity in order to achieve the requisite contrasts at planetary separations, their proper motions are invariably higher than those of distant background objects. Thus, most planet candidates are only confirmed after obtaining a second epoch observation months or years after the initial detection to confirm that the candidate and host star exhibit the same proper motions over that time period, as shown schematically in figure \ref{fig:propermotion}. Candidates are ruled out if they exhibit little to no proper motion between epochs.
 
In principle, establishment of common proper motion could be complicated by the additional motion of a true bound companion as it orbits its host star. In practice, however, most planet candidates are separated from their hosts by large enough physical separations that orbital motion is negligible compared to proper motion. 

The most insidious form of false positive in establishing common proper motion is the coincidental alignment of an unbound foreground or background object with non-negligible proper motion and the target star. If the proper motion vectors of the two object are in rough alignment and of similar magnitude, the time baseline needed to distinguish a comoving object is longer. Such was the case with the apparent planetary companion HD 131399Ab \citep{Nielsen2017}. 

\begin{figure}
    \centering
    \includegraphics[width=\textwidth]{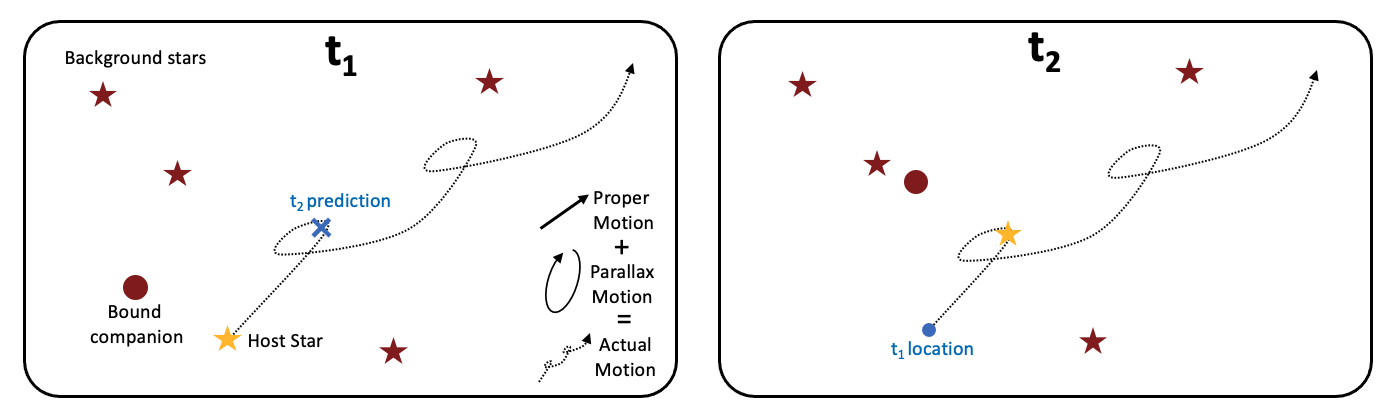}
    \caption{A schematic depiction of the process of determining common proper motion for a companion candidate (red circle) bound to a host star (yellow star). If the candidate is a true companion, then its motion over time (e.g. between epochs t$_1$ at left and t$_2$ at right) will closely follow the sky motion of the star (a combination of parallax and proper motion, shown as a dashed line). Companion host stars are generally close to Earth, with a higher degree of proper motion and parallax than more distant background stars, which move very little between epochs. The orbit of the bound companion around the host star (not depicted here) can complicate this somewhat, but orbital motion is generally slow for the widely-separated directly imaged companions detected to date. Importantly, color alone is rarely enough to determine whether a companion is bound or not, as background red giants share similar colors to directly imaged companions.}
    \label{fig:propermotion}
\end{figure}

\subsection{Disk Features}
Another form of astrophysical false positive results from the prevalence of circumstellar material around the young stars targeted for direct imaging. Upon PSF subtraction, disk features can masquerade as planets, especially in cases where they are narrow (surviving highpass filtering) and non-axisymmetric (not shared with many images in the reference library). This is especially problematic for younger systems ($<$10Myr), where such features are ubiquitous \citep[e.g.][]{Benisty2022}. 

In the case of older ($>$10Myr) objects, for which the initial protoplanetary disk has usually either been incorporated into companions or dissipated, we see primarily second generation dust generated by the grinding of asteroids and/or comets in belts akin to our own asteroid and Kupier belts. These belts tend to be fairly symmetric and have limited spatial extent, making them much less likely to be confused for planet candidates. In the case of known disk-bearing systems, candidate planets are vetted in several ways. 

\paragraph {Comparison with known disk features} in both millimeter thermal emission and NIR scattered light (especially PDI-resolved features) informs the probability of confusion occurring at the location of a planet candidate. In cases where a candidate is well inside of a cleared cavity \citep[e.g. PDS 70b,][]{Keppler2018}, the odds of confusion are minimal. 

\paragraph{Colors or spectra} of companion candidate(s) can be compared to those of the star. In a case where the star and candidate spectra closely match, odds are good that the candidate has a substantial scattered light component. This could mean an envelope or disk around a planet, or a clump of disk material that has not yet formed a planet.  In cases where a planet candidate exhibits a substantially different spectrum from that of the star, it is considered strong evidence for a planetary nature.

\paragraph{Multiepoch information} can be obtained to distinguish static disk features from orbiting companions. This is complicated in the case of disk features such as planet-induced spiral arms, which likely rotate with a pattern speed equal to the orbital speed of the companion inciting them. An important test is, therefore, whether apparent point sources that lie along spiral arms orbit with the speed of a companion at the point source's orbital separation. If they orbit faster (or slower), this is consistent with incitement by a different planet on a closer (or more distant) orbit. 

\paragraph{The robustness of the signal among post-processing techniques} particularly those that vary somewhat in ``aggressiveness". In the most insidious cases, the presence of an extended but narrow disk feature at different azimuths in the PSF reference library can lead it to appear point-like in post-processed images. Persistence of the feature across PSF subtraction algorithmic properties, and in particular its persistence across various HCI techniques, helps to distinguish this scenario from a true point-like source. In cases where the disk structures are well constrained (e.g. from PDI imaging), forward modeling can be used to understand the likely appearance of disk structures following PSF subtraction and compared against the images. RDI and cADI are considered the most conservative processing techniques, while LOCI-ADI and KLIP-ADI are more ``aggressive" in that they tend to model smaller spatial scale PSF features, and model mismatch can therefore result in smaller spatial scale apparent substructures that mimic planetary signals. Tunable parameters in the algortihms, such as the degree of rotational masking, the size of the regions for which PSFs are constructed separately, and the complexity/number of modes applied to construct the model, can be altered to be more or less aggressive. For example, including images in the reference library that are close in rotational space (a small rotational mask), constructing custom PSFs for very small regions of images, and increasing the number of modes in the PSF model all represent more ``aggressive" reductions that will effectively remove stellar signal, but will also increase the rate of false positives. These parameters can be relaxed or iterated over to probe the robustness of any apparent signals. 

Various other optical artifacts, quasi-static speckles, cosmic rays, and speckle noise can in principle masquerade as planets in post-processed images. In general, the properties of such artifacts should not closely mimic those of true astrophysical sources (e.g. by demonstrating self-subtraction). Nevertheless, careful analysis of false alarm probabilities is important in conducting HCI, particularly for low SNR recoveries. The gold standard in candidate vetting remains multiepoch, multi-wavelength, multi-instrument observations of candidates demonstrating common proper motion with the host star and evidence of a non-stellar spectrum.

\section{Other related technologies \label{sec:othertech}}

Although this tutorial is focused specifically on ground-based, non-interferometric direct imaging techniques, there are several highly related or complementary techniques that are worth highlighting. 

\paragraph{Interferometric Techniques} can be applied in HCI in several ways. First, the beams from multiple telescopes can be combined in the classic sense to both collect more light and achieve higher resolution than is achievable with a single telescope aperture (because the resolution of an interferometer is $\lambda$/2B, where B is the longest Baseline distance between telescopes). Even in the case where multiple telescopes are not available for use in the classical sense of an interferometer, a technique called ``Non-Redundant Aperture Masking" \citep{Nakajima1989} can be used to achieve higher resolution on a single telescope.  NRM requires the application of a pupil mask that is mostly opaque but contains a number of holes, each pair of which has a different separation and therefore probes a different spatial frequency. The maximum resolution achievable under this technique is half of the classical diffraction limit ($\lambda$/2B), giving a distinct advantage at tight inner working angles for imaging companions. All interferometric imaging requires some degree of image reconstruction and is innately model-dependent, but these techniques nevertheless open up additional discovery space at high spectral and/or spatial resolution.

\paragraph{Space-Based HCI} is another important complimentary technique. It shares many features with ground-based HCI, including the need for wavefront sensing and control, image post-processing, and application of differential imaging techniques. Adaptive optics is in principle unnecessary in space, though some space-based HCI concepts use much lower cadence active mirror control to correct for slower (e.g. thermal) drifts in the shape of incoming wavefronts. Reference differential imaging is in many ways more powerful in space because of the innate stability of space-based instrumental PSFs, allowing in some cases for a reference library composed of images of tens to hundreds of sources in addition to the science target. Although space-based telescopes cannot leverage the rotation of the Earth to accomplish Angular Differential Imaging, they can apply a similar technique called ``Roll Subtraction" by rotating the telescope around its optical axis during an imaging sequence. The amount of achievable rotation and the number of reference angles in such cases is small (e.g. 2 reference angles separated by $\sim$15deg), but has nevertheless proven effective at accomplishing differential imaging in space. Spectral Differential Imaging is more or less unchanged in the space-based imaging scenario, as is Polarized Differential Imaging  in principle, though there are no plans to include PDI capabilities on any near-future space-based HCI missions. 

\paragraph{Sub-mm Interferometry} is a fully unrelated technique to HCI, but is nevertheless highly complimentary, particularly for understanding scattered light disk features and protoplanets. Interferometric sub-mm arrays, particularly ALMA, provide a key piece of the puzzle in that they probe thermal emission from large grains in the midplane of disks. Together with information from NIR HCI of the surface layers of the disk, as well as millimeter emission from molecular gas species, a holistic picture of a disk system can be formed that encompasses all three key components - large grains, small grains, and gas. Very high-resolution millimeter continuum imaging can even probe the presence of circum\textit{planetary} dust and gas, compelling additional evidence for the presence of protoplanets. 

\section{Conclusion}

Over the past fifteen years, ground-based High-Contrast Imaging has proven to be a robust and versatile way to probe the properties of young exoplanets and circumstellar disks. Using adaptive optics and wavefront sensing/control algorithms, atmospheric scintillation can be sensed and corrected for, allowing large ground-based telescopes to achieve diffraction limited or nearly diffraction-limited imaging at optical and near infrared wavelengths. HCI instruments often utilize coronagraphy to apply first-order suppression of incoming starlight, allowing faint nearby signals to be detected. Differential imaging techniques are then applied to leverage polarimetric, spectroscopic, \rev{target object}, and angular diversity in the data to identify and remove starlight.  Post-processing algorithms with various degrees of complexity and aggressiveness are then applied to enable detection of signals that are several orders of magnitude fainter in contrast, as well as detailed spectroscopic, photometric, and astrometric characterization. Signals are vetted by demonstrating common proper motion with the host star, robustness to algorithmic parameters, consistency with forward models, diversity in polarimetric or spectral properties relative to their host stars, and/or persistence across epochs, wavelengths, and instruments. HCI instruments and reduction techniques are necessarily complex in order to overcome the tremendous contrast and angular resolution barriers required to directly isolate the light from exoplanets and circumstellar disks. Yet, these techniques provide the best future prospects for someday detecting and characterizing an exo-Earth. 

\rev{This tutorial was designed as an introduction for beginners, and is not comprehensive in its technical details. My hope is that it will enable those just getting started in the field to access more technical HCI instrument manuals and published results. To learn more about the current state of the art in high-contrast imaging, please see \url{bit.ly/beginHCI}, which provides a ``Reading/Viewing List for Beginning High-Contrast Imagers".}

\section{acknowledgements}
    I would like to thank the wonderful undergraduate and graduate students in my Spring, 2023 research group for the many group meeting sessions of figure critiques that they engaged in - this article is much better for their feedback. They are: Sarah Betti, Jada Louison, Cat Sarosi, Cailin Plunkett, Alyssa Cordero, and Adrian Friedman. Thank you to Kim Ward-Duong and Cat Sarosi for their thorough reviews of the text of the article, and to Bruce Macintosh, Mark Marley, Max Millar-Blanchaer, Ewan Douglas, Christian Marois, and Rob de Rosa for consulting on various parts of it. Thank you to the anonymous reviewer for their extremely constructive feedback, which greatly improved the article. Finally, a huge thank you to my team of ``internal" student reviewers - Giselle Hoermann, Kinsey Cronin, Jessica Labossiere, and Jingyi Zhang.




\end{document}